\documentclass[twocolumn]{aastex631}

\usepackage{ulem}
\usepackage{comment}
\usepackage[version=4]{mhchem}
\usepackage{amsmath}
\usepackage[caption=false]{subfig}
\usepackage{appendix}
\usepackage{here}
\usepackage[capitalize]{cleveref}

\newcommand{\diff}{\mathrm{d}}
\newcommand{\rp}{R_\mathrm{p}}
\newcommand{\rsur}{R_\mathrm{s}}
\newcommand{\mpl}{M_\mathrm{p}}
\newcommand{\mr}{M_\mathrm{r}}
\newcommand{\tint}{T_{\mathrm{int}}}
\newcommand{\del}{\mathrm{\partial}}
\newcommand{\ms}{M_{\mathrm{s}}}
\newcommand{\fxuv}{F_{\mathrm{XUV}}}
\newcommand{\rxuv}{R_{\mathrm{XUV}}}
\newcommand{\ktide}{K_{\mathrm{tide}}}
\newcommand{\mesc}{{\dot M_{\mathrm{esc}}}}

\newcommand{\tsur}{T_{\mathrm{s}}}

\newcommand{\teq}{T_{\mathrm{eq}}}

\newcommand{\cvol}{c_{\mathrm{v}}}

\newcommand{\lumint}{L_{\mathrm{int}}}

\newcommand{\tirr}{T_{\mathrm{irr}}}
\newcommand{\fintristic}{F_\mathrm{int}}

\newcommand{\xmh}{{X_{\mathrm{H}}}}
\newcommand{\xmhe}{{X_{\mathrm{He}}}}
\newcommand{\psurh}{p_{\mathrm{s, H}}}
\newcommand{\psurhe}{p_{\mathrm{s, He}}}
\newcommand{\tomhe}{T_{\mathrm{o, He}}}
\newcommand{\tomh}{T_{\mathrm{o, H}}}

\newcommand{\GK}{\Delta_{r_1} G^{\circ} (T)}
\newcommand{\GFeO}{\Delta_{\ce{FeO}} G^{\circ}(T)}
\newcommand{\GFe}{\Delta_{\ce{Fe}} G^{\circ}(T)}
\newcommand{\GO}{\Delta_{\ce{O2}} G^{\circ}(T)}
\newcommand{\fenvini}{(f_{\rm{env}})_{\mathrm{ini}}}
\newcommand{\wfeo}{(W_{\ce{FeO}})_{\mathrm{ini}}}

\newcommand{\erg}{\, \mathrm{erg}}

\newcommand{\kel}{\, \mathrm{K}}
\newcommand{\gram}{\, \mathrm{g}}

\newcommand{\pbar}{\, \mathrm{bar}}

\renewcommand{\d}{\ensuremath{\Delta}}

\newcommand{\lodot}{L_{\odot}}

\newcommand{\moplus}{M_{\oplus}}
\newcommand{\roplus}{R_{\oplus}}

\begin{document}

\title{Helium Depletion in Escaping Atmospheres of Sub-Neptunes: A Signature of Primary-to-Secondary Transition}

\correspondingauthor{Hiroyuki Kurokawa}
\email{hirokurokawa@g.ecc.u-tokyo.ac.jp}

\author{Issei Kobayashi}
\affiliation{Department of Earth and Planetary Sciences, Institute of Science Tokyo, 2-12-1 Ookayama, Meguro, Tokyo 152-8551, Japan}

\author[0000-0003-1965-1586]{Hiroyuki Kurokawa}
\affiliation{Department of General Systems Studies, The University of Tokyo, 3-8-1 Komaba, Meguro, Tokyo 153-8902, Japan}
\affiliation{Department of Earth and Planetary Science, The University of Tokyo, 1-1-1 Yayoi, Bunkyo, Tokyo 113-0033, Japan}

\author[0000-0003-2915-5025]{Laura Schaefer}
\affiliation{Department of Earth and Planetary Sciences, Stanford University, Stanford, CA 94305, USA}

\author[0000-0002-1886-0880]{Satoshi Okuzumi}
\affiliation{Department of Earth and Planetary Sciences, Institute of Science Tokyo, 2-12-1 Ookayama, Meguro, Tokyo 152-8551, Japan}



\begin{abstract}

Short-period sub-Neptunes are common in extrasolar systems. These sub-Neptunes are generally thought to have primary atmospheres of protoplanetary-disk gas origin. However, atmospheric escape followed by degassing from their interiors can lead to the transition to secondary atmospheres depleted in gases less-soluble to magma, such as helium. These primary and secondary atmospheres can potentially be distinguished from observations of escaping hydrogen and helium. This study aims to elucidate the impact of the primary-secondary transition on atmospheric compositions of short-period sub-Neptunes. We simulate their evolution with atmospheric escape driven by stellar X-ray and extreme ultraviolet irradiation and degassing of hydrogen, helium, and water from their rocky interiors, with a one-dimensional structure model. We show that the transition takes place for low-mass, close-in planets which experience extensive atmospheric escape. These planets show the depletion of helium and enrichment of water in their atmospheres, because of their low and high abundances in the planetary interiors, respectively. A compilation of our parameter survey (the orbital period, planetary mass, envelope mass, and mantle FeO content) shows a correlation between the planet radius and the helium escape rate. We suggest that the transition from primary to secondary atmospheres may serve an explanation for helium non-detection for relatively-small ($\lesssim 2.5\ R_\oplus$) exoplanets. 

\end{abstract}



\section{Introduction} \label{sec:intro}

Short-period sub-Neptunes, here defined as Earth- to Neptune-sized planets with orbital periods shorter than 100 days, are common in extrasolar systems. The number of confirmed exoplanets exceeds 6000, the majority of which are sub-Neptunes \citep{NEA}. Their radii and, when masses are estimated, bulk densities suggest that sub-Neptunes larger than 1.6 Earth radii typically possess thick atmospheres \citep{weiss+marcy2014,rogers2015most,fulton2017}, conventionally thought to be protoplanetary-disk-gas origin (hereafter called primary atmospheres).

Due to their proximity to their host stars, short-period sub-Neptunes experience extensive atmospheric escape. Intense stellar X-ray and extreme ultraviolet ($\lesssim 100\ \rm{nm}$ photons, hereafter called XUV) irradiation induces atmospheric escape \citep[hereafter referred as XUV-driven escape, e.g.,][]{murray2009atmospheric,lopez2012,owen&jackson2012,kurokawa2014mass,Malsky_2020,Koskinen_2022}, especially when stars are young and thus emit higher XUV flux due to their vigorous magnetic activities induced by faster rotation \citep{ribas2005}. Along with atmospheric boil-off \citep[e.g.,][]{Owen&Wu2016,Misener&Schlichting2021,rogers2023light}, core-powered mass-loss \citep[e.g.,][]{Ginzburg2018,Gupta&Schlichting}, and impact erosion \citep[e.g.,][]{Sakuraba+2019Icar..317...48S,Sakuraba+2021NatSR..1120894S,Izidoro2022,Krissansen-Totton+2024NatCo..15.8374K}, the XUV-driven escape is thought to have influenced atmospheric evolution of short-period sub-Neptunes.

Degassing of dissolved components from the magma ocean can replenish the atmosphere removed, extending the lifetime against the atmospheric mass-loss \citep{chachan2018,kite2020exoplanet}. 
High equilibrium temperature and the greenhouse effect of a thick atmosphere sustain a long-lived magma ocean on a short-period sub-Neptune, if the interior (the non-gaseous part) is primarily composed of rock. The magma ocean can contain a large amount of dissolved volatiles. Moreover, chemical reactions such as the formation of water can alter the atmospheric composition \citep{kite2020atmosphere,kite_schaefer2021,Dorn&Lichtenberg2021,Charnoz2023,Tian2023}.

Efforts to understand sub-Neptune atmospheres with transit spectroscopic observations are plentiful \citep[e.g.][]{tsiras2016,Wakeford2017,Benneke_2019}. Their spectra are often featureless, which prevents us from drawing firm conclusions \citep[e.g.][]{Knutson2014,chachan2020,Libby-Roberts_2020}, though recent JWST observations have revealed molecular features in some cases \citep{Kreidberg+Stevenson2025arXiv250700933K}. A widely-accepted explanation for their flat transmission spectra is the presence of aerosols including organic hazes generated by photochemical reactions \citep{kawashima2019theoretical} and mineral clouds \citep{ohno2021}. 

Other useful observables to probe atmospheres of transiting short-period exoplanets are escaping hydrogen and helium. 
The escaping atmosphere was first detected with atomic hydrogen absorption in the stellar Lyman-$\alpha$ line \citep{Vidal-Madjar+2003}. 
A significant advancement was recently achieved with the detection of helium in the extended atmosphere \citep{Spake+2018}. Helium in the extended atmosphere transitions to a metastable state known as the $2^{3}$S helium triplet through recombination or collisional excitation from the ground state \citep{Seager_2000,Oklop_i__2018}. An absorption line at $10830\ {\rm \AA}$ arises from resonant scattering of the $2^{3}$S state to the $2^{3}$P state. This line is located at the near-infrared wavelength region, enabling ground-based observations. 
Moreover, theoretical modeling predicts that the helium absorption is more sensitive to the escape rate than that of hydrogen \citep{Ballabio+Owen2025,Schreyer+2024}.

While models for the extended atmosphere of sub-Neptunes generally assume (proto)solar-like hydrogen to helium ratios \citep[e.g.,][]{Oklop_i__2018,Ballabio+Owen2025}, processes in planetary evolution can change their atmospheric composition. Preferential escape of hydrogen over helium with a moderate escape rate has been proposed to form a helium-rich atmosphere \citep{hu2015,cherubim2024Strong}. In contrast, intense atmospheric escape combined with degassing from the interior may lead to helium depletion, which is the possibility we consider in this study. The helium depletion, if it happened, may serve as an explanation for non-detection of escaping helium from some observed sub-Neptunes \citep[e.g.,][]{Kasper_2020,Vissapragada+2024ApJ...962L..19V}.

This study aims to elucidate the transition from primary to secondary atmospheres in the evolution of short-period sub-Neptunes. In addition to hydrogen and water, the latter of which forms from chemical reactions with the magma ocean, we explicitly consider helium as the third component, as it can serve as a proxy for the transition to the secondary atmosphere. We simulate the evolution with XUV-driven atmospheric escape and replenishment by degassing from the interior. We also investigate how the transition from the primary to the secondary atmosphere is correlated with other observables such as the planetary radius. We contextualize these results within the above-mentioned observations of escaping helium. We introduce our model in Section \ref{sec:methods}. Section \ref{sec:results} presents the results. Section \ref{sec:discussion} discusses implications for the observations of the $10830\ {\rm \AA}$ absorption line of the sub-Neptunes' atmospheres. We conclude in Section \ref{sec:conclusions}.

\section{Methods}\label{sec:methods}

Here we introduce our model for the planetary structure (Section \ref{subsec:methods:structure}), thermal evolution (Section \ref{subsec:methods:cooling}), atmospheric mass-loss (Section \ref{subsec:methods:escape}), chemical reactions to produce water (Section \ref{subsec:methods:reactions}), dissolution (Section \ref{subsec:methods:dissolution}), and fugacity (Section \ref{subsec:methods:fugacity}). Input parameters and the overall workflow is presented in Section \ref{subsec:methods:workflow}.

\subsection{Planetary structure}\label{subsec:methods:structure}

Our model for the planetary structure combines a spherically-symmetric, one-dimensional (1D) atmosphere \citep[e.g.,][]{rogers2011,Kurokawa+2013MNRAS.433.3239K,kurokawa2014mass} with a one-box (not spatially-resolved) interior.
Equations for the hydrostatic equilibrium, the mass conservation, the optical depth, and the state adapted for the atmosphere are given as,
\begin{equation}
    \frac{\diff \ln{p}}{\diff \mr} = -\frac{G\mr}{4\pi r^{4}p}\label{21},
\end{equation}
\begin{equation}
    \frac{\diff \ln{r}}{\diff \mr} = \frac{1}{4\pi r^{3}\rho}\label{22},
\end{equation}
\begin{equation}
    \frac{\diff \ln{\tau}}{\diff \mr} = -\frac{\kappa}{4\pi r^{2} \tau},
\end{equation}
\begin{equation}
    \rho = \rho\left(p,T,Y,Z\right),
\end{equation}
\begin{equation}
    \frac{\diff \ln{T}}{\diff M_{r}} = -\frac{GM_{r}}{4\pi r^{4}p} \cdot \min\left(\nabla_{\rm{rad}},\nabla_{\rm{ad}}\right),
\end{equation}
where $p$ is the pressure, $r$ is the distance from the planet center, $\mr$ is the enclosed mass, $T$ is the temperature, $\rho$ is the mass density, $\tau$ is the optical depth, $Y$ and $Z$ are the mass fractions of helium and water, $\kappa$ is the Rosseland mean opacity for thermal radiation, $G$ is the gravitational constant, $\nabla_{\rm{rad}}$ is the temperature gradient of the diffusion approximation for radiation, and $\nabla_{\rm{ad}}$ is the adiabatic temperature gradient.

The Rosseland mean opacity for the hydrogen and helium mixture was obtained from \cite{freedman2014}.
We used the Rosseland mean opacity data table for water from \cite{KurosakiIkoma2017}. In addition to the opacities for thermal radiation, our model for the stratospheric temperature profile (Equation \ref{eq:str1}) also requires opacities for stellar (visible) radiation, which were likewise taken from these studies.
For simplicity, we calculated the Rosseland mean opacity of the hydrogen/helium and water mixture as their mass-weighted average.

We used the equation of state (EoS) for non-ideal hydrogen and helium gas given by \cite{saumon1995}. For water, we used the EoS given by \cite{Haldemann_2020}.
Using the EoSs for hydrogen, helium, and water, we determined the density of the mixture as a function of temperature, pressure, and composition. The density is given as,
\begin{equation}
    \frac{1}{\rho} = \frac{1 - Y - Z}{\rho_{\ce{H}}} + \frac{Y}{\rho_{\ce{He}}} + \frac{Z}{\rho_{\ce{H2O}}}, 
\end{equation}
where $\rho_{\ce{H}}$, $\rho_{\ce{He}}$, and $\rho_{\ce{H2O}}$, are the density of hydrogen, helium, and water at a given set of temperature and pressure.

The atmosphere consists of upper radiative-equilibrium and lower convective layers: the stratosphere and the troposphere. In the upper atmosphere, we assumed a radiative-equilibrium profile with the two-stream approximation \citep{guillot2010}.
\begin{equation}
    \label{eq:str1}
    \begin{split}
    T^{4} &= \frac{ \tint^{4}}{4}\left[\frac{2}{3} + \tau
    \right] \\ 
    &\quad + \frac{3\tirr^4}{4}f \left[\frac{2}{3} + \frac{1}{\gamma\sqrt{3}} + \left(\frac{\gamma}{\sqrt{3}} - \frac{1}{\gamma \sqrt{3}}\right) e^{-\gamma\tau\sqrt{3}}\right],
    \end{split}
\end{equation}
where $\tint$ is the intrinsic temperature, and $\tirr$ is the irradiation temperature.
The irradiation temperature characterizes the irradiation from the central star. The relation between the irradiation temperature and the equilibrium temperature is expressed as $\teq = f^{1/4}\tirr$, where $f$ is the redistribution coefficient for the energy and is assumed to be $1/4$ in this study (full redistribution).
We calculated $\gamma$ by taking the ratio of visible to thermal (infrared) Rosseland mean opacities. 
Each opacity is obtained from the aforementioned tables.
In the optically thick region of $\tau \gg 1/({\sqrt{3}\gamma})$, we adopted the smaller temperature gradient between that of the radiation diffusion approximation and the adiabatic temperature gradient, given as,
\begin{equation}
\label{eq:str2}
    \nabla_{\rm{rad}} = \frac{3 \kappa p r^{2}}{16 \sigma_\mathrm{SB} T^{4} G \mr } \cdot
    \frac{\lumint}{4 \pi r^{2} } ,
\end{equation}
\begin{equation}
    \label{eq:tr}
    \nabla_{\rm{ad}} = \left(\frac{\del \ln{T}}{\del \ln{p}}\right)_{S} ,
\end{equation}
where $\lumint$ is the intrinsic luminosity of the planet and $\sigma_\mathrm{SB}$ denotes the Stefan–Boltzmann constant.
We obtained the entropy of the mixture by adding those of hydrogen, helium, and water from the EoS tables \citep{saumon1995,Haldemann_2020} and the mixing entropy calculated with a method summarized by \cite{Baraffe2008}. Then the adiabatic gradient was calculated with a method summarized by \cite{saumon1995}.

We defined the upper boundary of the atmosphere as the layer where $\tau = 2/3$.
At the upper boundary, the boundary conditions are given as $r = \rp$, $M_r = \mpl$, $T = T_{\rm{ph}}$, and $p = p_{\rm{ph}}$, where $\rp$ is the planet radii, and $\mpl$ is the planet mass.
The temperature $T_{\rm{ph}}$ is given by Equation (\ref{eq:str1}) with $\tirr$ and $\tint$. The irradiation temperature $\tirr$ is given by $\tirr = f^{-1/4} \teq$. The equilibrium temperature $\teq$ is expressed as the function of the distance from the host star 
\begin{equation}
    \label{Teq}
    \sigma_\mathrm{SB} \teq ^ {4} = \left( \frac{L_*}{16 \pi d^{2}} \right)  \left(1-A\right),
\end{equation}
where $d$ is the distance from the host star, $L_*$ is the stellar luminosity, and $A$ is the albedo. In this study, we fixed $A = 0$ and $L_*=\lodot=3.826 \times 10^{33}\ \mathrm{erg \, s^{-1}}$. The intrinsic temperature $\tint$ is given by $\tint = (\lumint/4\pi\rp\sigma_\mathrm{SB})^{1/4}$. The pressure at the upper boundary $p_{\rm{ph}}$ is given from an approximation using the equation of the hydrostatic equilibrium:
\begin{equation}
    p_{\rm{ph}} \simeq \frac{g_{\rm{ph}}\tau_{\rm{ph}}}{\kappa_{\rm{ph}}}.
\end{equation}
where $\kappa_{\rm{ph}}$ is the Rosseland mean opacity at the photosphere, and $g_{\rm{ph}} = {GM_{p}}/{R_{p}^2}$ is the gravity at the photosphere, and $\tau_{\rm{ph}}/\kappa_{\rm{ph}}$ gives the mass column density.

For the interior model, we utilized the fitting relation for the mass and radius for rocky planets with an Earth-like composition, given as \citep{valencia2006internal,kite2020atmosphere},
\begin{equation}
    \label{underbound}
    {R_{\rm{s}}} = \left(\frac{M_{\rm{s}}}{M_{\oplus}}
    \right)^{0.27}{\roplus},
\end{equation}
where $\rsur$ is the radius of the rocky surface, $\ms$ is its enclosed mass, $M_{\oplus}$ is the Earth mass, and $\roplus$ is the Earth radius. This model assumes a solid interior, but a liquid silicate interior likely makes a difference of only a few percent or less in planetary radius \citep{Boley+2023ApJ...954..202B}.

\subsection{Thermal evolution} \label{subsec:methods:cooling}

For thermal evolution, we integrated the equation for the energy balance \citep[e.g.,][]{lopez2012,kurokawa2014mass},
\begin{equation}
    \label{eq:thermal}
    \int_{\ms}^{\mpl} \diff \mr \frac{T\diff S}{\diff t} = -4\pi \rp ^{2} \fintristic -c_{\mathrm{v}} \ms \frac{\diff T_{\mathrm{s}}}{\diff t}, 
\end{equation}
where $\cvol$, $\tsur$, and $S$ are the specific heat capacity of the rock, the temperature at the magma--atmosphere boundary, and the entropy, respectively. The left-hand side of Equation (\ref{eq:thermal}) is the time derivative of the total energy of the envelope. The first term on the right side conforms to the total radiation heat flux of the planet. The second term on the right side corresponds to the amount of energy decrease per time in the rocky part of the planet.
We assume $\cvol = 10^{7}\ \erg  \gram^{-1} \kel^{-1}$ \citep[e.g.,][]{guillot1995,lopez2012,chachan2018}.

We approximated the whole atmosphere as isentropic when solving Equation (\ref{eq:thermal}) for the sake of simplicity. This assumption can be applied accurately in the troposphere but breaks down above it. However, this simplification does not affect the outcome since the majority of the envelope mass is in the troposphere.

\subsection{Atmospheric escape} \label{subsec:methods:escape}

The XUV flux from the central star heats the upper atmosphere, which causes atmospheric escape. We adopted the mass-loss rate of the energy-limited atmospheric escape given as \citep{1981Watson, Erkaev2007},
\begin{equation}
    \label{eq:Mesc}
    \mesc = \frac{\eta \pi \fxuv \rxuv ^{3}}{G \mpl \ktide},
\end{equation}
where $\rxuv$ is the radius at which the optical thickness of XUV is unity. We assumed $\rxuv = \rp$ for the simplicity. In reality, XUV photons are absorbed in the upper atmosphere. In our nominal case ($M_\mathrm{p} = 15\ M_\oplus$, $R_\mathrm{p} \simeq 3\ R_\oplus$, and $T_\mathrm{eq} \simeq 10^3\ \mathrm{K}$), we estimate $R_\mathrm{XUV} \simeq 20\times H_R \simeq 1.1\times R_\mathrm{p}$, where $H_R$ is the scale height at $r=R_\mathrm{p}$ \citep{murray2009atmospheric}. The rate of energy reduction due to tidal forces from stars, $\ktide$, is a dimensionless parameter and we fixed $\ktide = 1$ following \cite{chachan2018}. The efficiency of conversion of XUV flux energy to atmospheric escape (evaporation efficiency), $\eta$, is fixed at 0.1. Hydrodynamic modeling indicates that the efficiency $\epsilon$ varies, but typically stays within a factor of a few around $0.1$ in our parameter range of interest \citep{owen&jackson2012,Caldiroli+2021A&A...655A..30C,Caldiroli+2022A&A...663A.122C}. As our purpose is to show how the primary-secondary transition impacts atmospheric compositions of short-period sub-Neptunes, we consider this simplification to be justified.\par
The time evolution of XUV flux is given by \citep{ribas2005},
\begin{equation}
    \label{eq:xuv}
    \fxuv = 2.97 \times 10^{-2} \left(
    \frac{t}{1\,\mathrm{Gyr}}
    \right)^{-1.23}
    \left(
    {\frac{d}{1\,\mathrm{au}}}
    \right)^{-2} \mathrm{\,W\, m^ {-2}},
\end{equation}
where  $\fxuv$ and $t$ are the XUV flux from the star and time, respectively. \Cref{eq:xuv} corresponds to the XUV flux from a sun-like star. 
We assumed the saturation phase of $t < 100\ \mathrm{Myr}$; the XUV flux in this period is fixed at the value for $t = 100\ \mathrm{Myr}$ in \Cref{eq:xuv} \citep{scalo2007}.

We assumed that atmospheric escape induces no elemental fractionation. In reality, atmospheric escape preferentially removes lower-molecular-weight species for a lower mass-loss rate \citep{catling&kasting2017}. Significant helium enrichment has been predicted for planets whose orbital periods $\gtrsim 5 \ {\rm days}$ \citep{cherubim2024Strong}. Our assumption is justified because we primarily focus on shorter-period planets: $d = 0.025$ to $0.075\,\mathrm{au}$, which corresponds to $1.4$ to $7.5\ {\rm days}$ (\Cref{table: parameters}). We will address the effects of elemental fractionation due to atmospheric escape in Section \ref{subsec:discussion:fractionation}.

\subsection{Production of water} \label{subsec:methods:reactions}

In addition to hydrogen and helium accreted from protoplanetary disk gas, we considered water formed with chemical reactions of hydrogen with iron oxide in magma. This is because water, highly soluble to magma, may act as another proxy for the secondary atmosphere, and because production of water changes atmospheric composition, its scale height, planet size, and consequently, its evolution. Our model follows that of \cite{kite2020atmosphere}. A key reaction is,
\begin{equation}
    \label{reaction:H2FeO}
    \ce{H2} +\ce{FeO} = \ce{H2O} + \ce{Fe}.
\end{equation}
The amount of \ce{H2O} produced is determined by finding the equilibrium state of this equation using the amount of \ce{H2} initially added and the amount of \ce{FeO} present in the magma ocean.

To find the equilibrium state of Reaction (\ref{reaction:H2FeO}), we divided this reaction into two parts. The first reaction is,
\begin{equation}
    \label{reaction:FeO2}
    \ce{2Fe} + \ce{O2} = \ce{2FeO}.
\end{equation}
The equilibrium constant for Reaction (\ref{reaction:FeO2}) is,
\begin{equation}
    \label{K1}
    K_{\rm eq,1} = \frac{[\ce{FeO}]^2}{[\ce{Fe}]\frac{f\ce{O2}}{f\ce{O2}^{\circ}}},
\end{equation}
where the molecular formulae in brackets are the chemical activities, and $f\ce{O2}$, $f\ce{O2}^{\circ}$, and $K_{\rm eq,1}$ are the oxygen fugacity, the oxygen fugacity at the standard pressure, and the equilibrium constant.
We assume $[\ce{Fe}] = 1$ because iron is considered to exist as immiscible droplets in the magma ocean. Furthermore, the activity coefficient of \ce{FeO} is fixed at 1.5 \citep{HOLZHEID1997}. A chemical activity is given by the product of the activity coefficient and the molar fraction.
The following thermodynamic relation holds between the chemical equilibrium constant and the standard reaction Gibbs energy,
\begin{equation}
    \label{K_G_relation}
    K_{\rm eq} = \exp{\left(\frac{-\Delta G}{RT}\right)},
\end{equation}
where $R$ is the molar gas constant and $\Delta G$ is the Gibbs energy of reaction.
Using \Cref{K_G_relation}, the chemical equilibrium constant can be obtained from the standard state Gibbs energy of reaction.
The standard state Gibbs energy of Reaction (\ref{reaction:FeO2}) is given by,
\begin{equation}
    \GK = 2\GFeO - 2\GFe - \GO,
\end{equation}
where $\GK$, $\GFeO$, $\GFe$, $\GO$ are the standard state Gibbs energy of Reaction (\ref{reaction:FeO2}), and the standard state Gibbs energy of formation at a given temperature of \ce{FeO}, \ce{Fe}, and \ce{O2}.
Each standard Gibbs energy for formation was obtained from \cite{KOWALSKI1995229} as follows,
\begin{equation}
    \label{equation:GFeO}
    \begin{split}
        \mathrm{GFEO} &= -279318 + 252.848 T -46.12826\ T \log{T} \\
        & - 0.0057402984\ T^{2} \ (298.15 \kel < T < 3000 \kel),\\
    \end{split}
\end{equation}
\begin{equation}
    \GFeO = \mathrm{GFEO} + 34008 - 20.969 T,
\end{equation}
\begin{multline}
    \mathrm{GHSERFE} = 1225.7 + 124.134\ T \\ + 23.5143\ T \log{T} - 0.00439752\ T ^ {2} \\
    - 0.0000000589269\ T^{3} + 77358.5 / T,\\
\end{multline}
\begin{equation}
    \begin{split}
        \GFe = 
        \begin{cases}
             12040.17 - 6.55843\ T - 3.6751551\ T^{7} \times 10^{-21}\\
             + \mathrm{GHSERFE} &\\
             \quad  (T < 1811 \kel),\\
             -10839.7 + 291.302\ T - 46\ T \log{T} &\\
             \quad (1811 \kel \leq T < 6000 \kel),
        \end{cases}
    \end{split}
\end{equation}
\begin{equation}
    \begin{split}
        \GO = 
        \begin{cases}
        -6961.744 - 76729.7484 / T - 51.0057202\ T\\ -22.2710136
         T  \log{T} - 0.0101977469\ T ^ {2}\\ + 1.3236921\ T^ {3} \times 10^{-6} &\\
         \quad (T < 1000 \kel),\\
        -13137.5203 + 525809.556/T + 25.3200332\ T\\ - 33.627603\ T \log{T}
        -0.00119159274\ T^ {2}\\ + 1.3561111\ T ^{3} \times 10^{-8} &\\ 
        \quad (1000 \kel \leq T < 3300 \kel),\\
        - 27973.4908 + 8766421.4\ / T + 62.5195726\ T \\
        - 37.9072074\ T \log{T} 
        - 8.5048377\ T ^ {2} \times 10^{-4} \\ + 2.1440978\ T ^ {3} \times 10^{-8}&\\
        \quad( 3300 \kel \leq T < 6000 \kel).
        \end{cases}
    \end{split}
    \label{eq:GO2}
\end{equation}
Since this study targets planets with thick atmospheres in the vicinity of stars, the magma-atmosphere boundary temperature of the planets can be as high as several thousand kelvins.
In addition, immediately after planetary formation, the planet has a very high temperature such as 7000 K due to the release of gravitational energy from the accreted mass.
Therefore, the Gibbs free energy for those high temperature was obtained by extrapolation with Equations (\ref{equation:GFeO})--(\ref{eq:GO2}).

To obtain the Gibbs energy at a given pressure from the standard state Gibbs energy, we considered the basic thermodynamic equation:
\begin{equation}
    \Delta_\mathrm{{r_1}} G = \GK + \int_{p_\mathrm{ref}}^p \Delta V dp, \label{eq:dG}
\end{equation}
where we assume a standard state reference pressure of 1 bar. To determine the pressure dependence of the Gibbs energy for \ce{FeO}, we utilized the Tait equation of state. The pressure-dependent term in Equation (\ref{eq:dG}) is given by the following expression \citep{Armstrong2019}:
\begin{equation}
    \int_{1\rm{bar}}^p Vdp = p V_{0, T} 
    \left[1 - a + \frac{a \{1 - (1 + bp)^{(1-c)} \} }{b(c-1)p}
    \right].
\end{equation}
Here $a$, $b$, and $c$ are defined as follows:
\begin{align}
    \begin{split}
        a &= \frac{1 + K_{0}^{\prime}}{1 + K_{0}^{\prime} + K_{0} K_{0}^{\prime \prime}}, \\
        b &= \frac{K_{0}^{\prime}}{K_{0}} - \frac{K_{0}^{\prime \prime}}{1 + K_{0}^{\prime}}, \\
        c &= \frac{1 + K_{0}^{\prime} + K_{0}^{\prime}K_{0}^{\prime 2}}{K_{0}K_{0}^{\prime 2} + K_{0}^{\prime} - K_{0}K_{0}^{\prime \prime}}, 
    \end{split}
\end{align}
where $K_{0}$, $K_{0}^{\prime}$, $K_{0}^{\prime \prime}$ are the bulk modulus at ambient conditions, the first pressure derivative, and the second pressure derivative. The values of the bulk modulus and its derivatives for FeO are, $K_{0} = 45.8\ {\rm GPa}, \, K_{0}' = 4.7$, and $K_{0}'' = - K_{0}' / K_{0}$.
Volume at the ambient pressure $V_{0, T}$ is given by $V_{0, T} = 13650 + 2.92 ( T - 1673 \kel )\, \rm{J/GPa}$.

We obtained the pressure-dependent term of Gibbs energy for \ce{Fe} from \cite{Komabayashi2014}. The Rose--Vinet EoS at the ambient pressure and temperature is given by,
\begin{equation}
    \label{eq:Rose-Vinet}
    p_{298} = 3 K_{0} x^{-2} (1-x) \exp{\left[\frac{3}{2} (K' - 1) (1 - x) \right]},
\end{equation}
where $x \equiv (V/V_{0})^{-1/3}$, $p_{298}$ is the pressure at $T = 298\ {\rm K}$, and $V_{0}$ is the molar volume at ambient pressure and temperature, respectively. Their values for Fe are, $K_{0} = 148 \, \mathrm{GPa}$, $K' = 5.8$, and $V_{0} = 6.88 \, \mathrm{cm^{3}\, mol^{-1}}$.
For the application to high-pressure conditions, the following equation is employed:
\begin{equation}
    \label{eq:VFE_T}
    V_{T} = V \exp{[ \alpha (T - 298)]}.
\end{equation}
Here, the following equation holds:
\begin{equation}
    \frac{\alpha}{\alpha_{0}} = \exp{\left[- \frac{\delta_{0}}{k} \left(
    1 - \frac{V}{V_{0}}^{k}\right) 
    \right]}
\end{equation}
where $\alpha_{0} = 9 \times 10^{-5} \, \kel^{-1}$, $\delta_{0} = 5.1 \, $, and $k =0.56$ are the thermal expansion coefficient at the normal temperature and pressure, the Anderson-Grüneisen parameter at the standard-state pressure, and the bulk compressibility at the standard temperature and pressure, respectively.
By employing the Newton-Raphson method to Equation (\ref{eq:Rose-Vinet}) and calculating temperature-dependence from Equation (\ref{eq:VFE_T}), we can determine the volume $V$ at any given temperature and pressure. Subsequently, through integration, we can obtain the Gibbs energy for liquid iron.  The equation of the reaction Gibbs energy for Reaction (\ref{reaction:FeO2}) is given by,
\begin{equation}
    \begin{split}
    \Delta_{\rm{r1}} G &= \GK + 2\int_{1\rm{bar}}^p V_{\ce{FeO}}dp \\ &\quad -  2\int_{1\rm{bar}}^p V_{\ce{Fe}}dp - RT \ln{\frac{f_{\ce{O2}}}{f_{\ce{O2}}^{\circ}}}.
    \end{split}
\end{equation}
The obtained Gibbs energies for Fe and FeO, along with Equations (\ref{K1}) and (\ref{K_G_relation}), allow us to determine the oxygen fugacity.

The second chemical reaction which connects Reactions (\ref{reaction:H2FeO}) and (\ref{reaction:FeO2}) is, 
\begin{equation}
    \label{reaction:H2O2_H2O}
    2\ce{H2} + \ce{O2} = 2\ce{H2O}.
\end{equation}
The equilibrium constant for Reaction (\ref{reaction:H2O2_H2O}) is,
\begin{equation}
\label{eq:K2}
    K_{\rm eq,2} = \left( \frac{f_{\ce{H2O}}/f_{\ce{H2O}}^{\circ}}{f_{\ce{H2}}/f_{\ce{H2}}^{\circ}}
    \right)^2 
    \frac{1}{f_{\ce{O2}}/f_{\ce{O2}}^{\circ}},
\end{equation}
where the standard state fugacities $f_{\ce{H2O}}$, $f_{\ce{H2}}$, and $f_{\ce{O2}}$ are all equal to 1 bar. Similarly, considering Equation (\ref{K_G_relation}) for $K_{\rm eq,2}$, the ratio of fugacities can be expressed by the following equation:
\begin{equation}
    \frac{f_{\ce{H2O}}}{f_{\ce{H2}}} = {\left(\frac{f_{\ce{O2}}}{f_{\ce{O2}}^{\circ}} \right)}^{0.5}\exp {\left(- \frac{\Delta G^{\circ} (T) }
    {4RT}\right)
    },
\end{equation}
where,
\begin{multline}
    \Delta G^{\circ} (T) = -4.8716 \times 10^{5} + 94.261574\ T + \\ 
    9.9275922\ T^{2} - 1.87633188 \times 10^{-6}\ T ^ 3 + 1.2446526\times 10^{-10} \times T ^ 4
\end{multline}
is the Gibbs energy at the standard pressure for \Cref{reaction:H2O2_H2O}.

\subsection{Dissolution} \label{subsec:methods:dissolution}

Water generated through the chemical reaction between hydrogen and iron oxide dissolves into the magma ocean. The solubility of \ce{H2O} is \citep{Schaefer_2016},
\begin{equation}
\label{eq: water dissolution}
    X_{\ce{H2O}} = 3.44 \times 10^{-8} \left(
    \frac{p_{\ce{H2O}}}{\rm{1 Pa}}\right) ^ {0.74}, 
\end{equation}
where $X_{\ce{H2O}}$ and $p_{{\ce{H2O}}}$ are the mass fraction of the \ce{H2O} in the magma ocean and the partial pressure of the \ce{H2O}. This dissolution model is based the dissolution data of basaltic magma \citep{Papale1997}. However, \cite{Sossi+2023E&PSL.60117894S} recently reported that the solubility of water in peridotite magma differs from that in basaltic magma by only $\simeq$10--20\%.

Hydrogen and helium dissolve into the magma ocean as well. The solubility of hydrogen, modified after \cite{chachan2018}, is given by,
\begin{equation}
    \label{eq: H2dissolution}
    \xmh = A_{\ce{H}} \cdot \psurh \cdot e^{- \tomh / \tsur},
\end{equation}
where $\xmh$, $\tsur$, and $\psurh$ are the solubility of hydrogen, the temperature at the boundary between magma and atmosphere, and partial hydrogen pressure at the magma-atmosphere boundary, and $\tomh$ and $A$ are constants.
The constant, $\tomh$ expresses the repulsive interaction of the molecule with the magma. We fixed $\tomh = 3000 \kel$. The other constant $A_{\ce{H}}$ satisfies $\xmh = 0.001\times 1/5$ when $\psurh = 1.5\, \mathrm{k}\pbar$, $\tsur = 1673 \kel$ based on the compiled data given by \cite{hirschmann2012}.

The factor $1/5$ is the correction term to convert the solubility in basaltic magma to the solubility in mantle peridotite. \cite{chachan2018} used hydrogen solubility in molten basalt to calculate the amount of hydrogen dissolved in magma oceans. Later, \cite{kite2020exoplanet} pointed out that hydrogen solubility in the mantle peridotite is about five times smaller than the basalt. As the peridotite composes Earth's mantle, we considered that it is appropriate to assume sub-Neptunes' magma ocean being composed of peridotite as well. Our model simply considers hydrogen solubility in peridotite by multiplying the solubility equation used in \cite{chachan2018} by $1/5$. 
We did not use the fugacity-dependent solubility equation used in \cite{kite2020atmosphere} to keep consistency with the assumed water solubility (Equation (\ref{eq: water dissolution})) being a pressure-dependent form. 

The solubility of helium in the magma ocean is given by,
\begin{equation}
    \label{eq: Hedissolution}
    \xmhe = A_{\ce{He}}\cdot \psurhe \cdot e^{- {\tomhe} / \tsur},
\end{equation}
where $\xmh$ and $\psurhe$ are the solubility of helium and the partial helium pressure at the magma-ocean boundary.
This equation was obtained by fitting experimental data at a high-temperature region \citep{IACONOMARZIANO2010}. We fixed $\tomhe = 4080 \, \kel$, which is obtained by assuming $\ln{\xmhe} \propto 1/\tsur$ in the fitting equation. We chose the constant parameter $A_{\ce{He}}$ to ensure that helium solubility per pressure is $e^{-14.48} \, \mathrm{Mpa}^{-1}$ at a magma-atmosphere boundary temperature $\tsur = 1673  \kel$.

\subsection{Fugacity}
\label{subsec:methods:fugacity}

The determination of partial pressures of \ce{H2} and \ce{H2O} requires obtaining the values of fugacity. The fugacity of species $i$ is given by,
\begin{equation}
    f_{i} = \phi_{i} p_{i},
\end{equation}
where $p_i$ is the partial pressure of species $i$, and $\phi_{i}$ is the fugacity coefficient defined as,
\begin{equation}
    \ln{\phi} = \int_{0}^{p} \frac{Z_{\rm{cf}} - 1}{p} dp,
\end{equation}
where $Z_{\rm{cf}}$ is the compressibility factor given by,
\begin{equation}
    Z_{\rm{cf}} = \frac{pV_{\rm{m}}}{RT} = \frac{m p}{\rho R T},
\end{equation}
where $V_{\rm{m}}$ and $m$ are the molar volume and the mean molecular weight, respectively.  The compressibility factor represents the deviation of real gases from ideal gas behavior, for an ideal gas, $Z_{\rm{cf}} = 1$. Following \cite{kite2020atmosphere}, we assumed $Z_{\rm{cf}} = 1$  for $Z_{\rm{cf}} < 1$ region to remove the effects of \ce{H2} and \ce{H2O} dissociation.
We calculated the fugacity coefficients of hydrogen and \ce{H2O} from their EoS \citep{saumon1995,Haldemann_2020}.

\subsection{Input parameters and workflow}
\label{subsec:methods:workflow}

\begin{table*}[h]
\centering
\setlength{\tabcolsep}{10pt} 
\renewcommand{\arraystretch}{1.0} 
\begin{tabular}{l|llll}
\hline \hline
Parameter                           & \multicolumn{4}{l}{Discrete values used in models}                   \\ \hline
$M_{\mathrm{p}}$                    & 5 $M_{\oplus}$ & 10 $M_{\oplus}$ & 15 $M_{\oplus}$$^{\dagger}$ & 25 $M_{\oplus}$ \\[0.5ex]
$d$                                 & 0.025 au$^{\dagger}$       & 0.05 au         & 0.075 au        &                 \\[0.5ex]
$(f_{\rm{env}})_{\rm ini}$                  &    1 wt\%$^{\dagger}$      &    3 wt\%      &    5 wt\%                &                 \\[0.5ex]
$(W_{\ce{FeO}})_{\mathrm{ini}}$     & 0 wt\%         & 8.24 wt\%$^{\dagger}$       & 49 wt\%         &                 \\[0.5ex]
$T_{\mathrm{s}}$                  & 5000 K         & 7000 K$^{\dagger}$          &                 &                 \\[0.5ex]
Reaction scenario & \multicolumn{4}{l}{Scenarios A$^{\dagger}$ and B; see the main text} \\ \hline
\end{tabular}
\caption{The input parameters and their corresponding values used in this study. The superscript $^\dagger$ indicates the nominal setting.}
\label{table: parameters}
\end{table*}

\subsubsection{Input parameters} \label{ss:input}

The input parameters of this study are the initial total mass of planets $\mpl$, the semi-major axis $d$, the initial envelope mass fraction $(f_{\rm{env}})_{\rm ini}$, the initial \ce{FeO} mass fraction in the magma ocean $(W_{\ce{FeO}})_{\mathrm{ini}}$, and the initial magma-atmosphere boundary temperature $T_{\mathrm{s}}$ (\cref{table: parameters}).
Regarding the initial molecular composition of the atmosphere, we considered two endmember scenarios for disk gas-atmosphere exchange during the formation stage (Scenarios A and B, see below).
We assumed $(W_{\ce{FeO}})_{\mathrm{ini}} = 8.24\, \mathrm{wt\%}$, $\fenvini = 1 \, \mathrm{wt}\%$, $T_{\mathrm{s}} = 7000 \kel$, $\mpl = 15\ \moplus$, $d = 0.025\, \mathrm{au}$, and Scenario A as nominal parameters.

The orbital distances considered in this study (0.025 au, 0.05 au, and 0.075 au from a sun-like host star) correspond particularly to sub-Neptunes located close to their host stars. 
Planets in such proximity to their stars are exposed to very intense stellar irradiation, receiving stellar flux approximately 100--2000 times greater than Earth's. 
The assumed range of semi-major axis corresponds to existing sub-Neptunes for which helium escape has been detected.

Regarding the planetary mass, we conducted the calculations for $5\ \moplus$, $10\ \moplus$, $15\ \moplus$, and $25\ \moplus$. 
This range covers those of sub-Neptunes. 
As observations of escaping helium have also been conducted for higher-mass planets, we also included $25\ \moplus$ in the parameter survey.

We considered 1 wt\%, 3 wt\%, and 5 wt\% of the planetary mass as the initial atmospheric amounts. 
For observed sub-Neptunes, the current envelope mass fraction is suggested to vary between 0.1--10\% \citep{lopez2014}. 
Our choice is typical for the atmospheres of sub-Neptunes and, as we show in Section \ref{sec:results}, appropriate to study the transition from the primary to secondary atmospheres.

We conducted simulations for three values of $(W_{\ce{FeO}})_{\mathrm{ini}}$: 0 wt\%, 8.24 wt\%, and 49 wt\%. 
The no \ce{FeO} case was considered as a reference model without water production.
The latter two cases correspond to the Earth-like mantle \ce{FeO} content and the case where complete oxidation of Earth's mantle+core occur \citep{kite2020atmosphere}.

Given the high uncertainty in initial planetary magma-atmosphere boundary temperatures, two different temperatures were employed in this study. 
Planets with higher initial temperatures cool down more rapidly, eventually reaching the same luminosity regardless of their initial temperature \citep[e.g.,][]{Marley+2007,kurokawa2014mass,Owen+Lai2018}. 
Thus, inferring the initial temperature from observations of an evolved exoplanet is challenging. 
This study set initial magma-atmosphere boundary temperatures at 5000 K and 7000 K for computation to elucidate the impact of initial thermal content on evolution. 
However, we show that the initial temperature does not significantly influence the evolution of a sub-Neptune's atmosphere (Section \ref{subsec:results:temperature}).

\begin{figure}
    \centering
    \includegraphics[width=\linewidth]{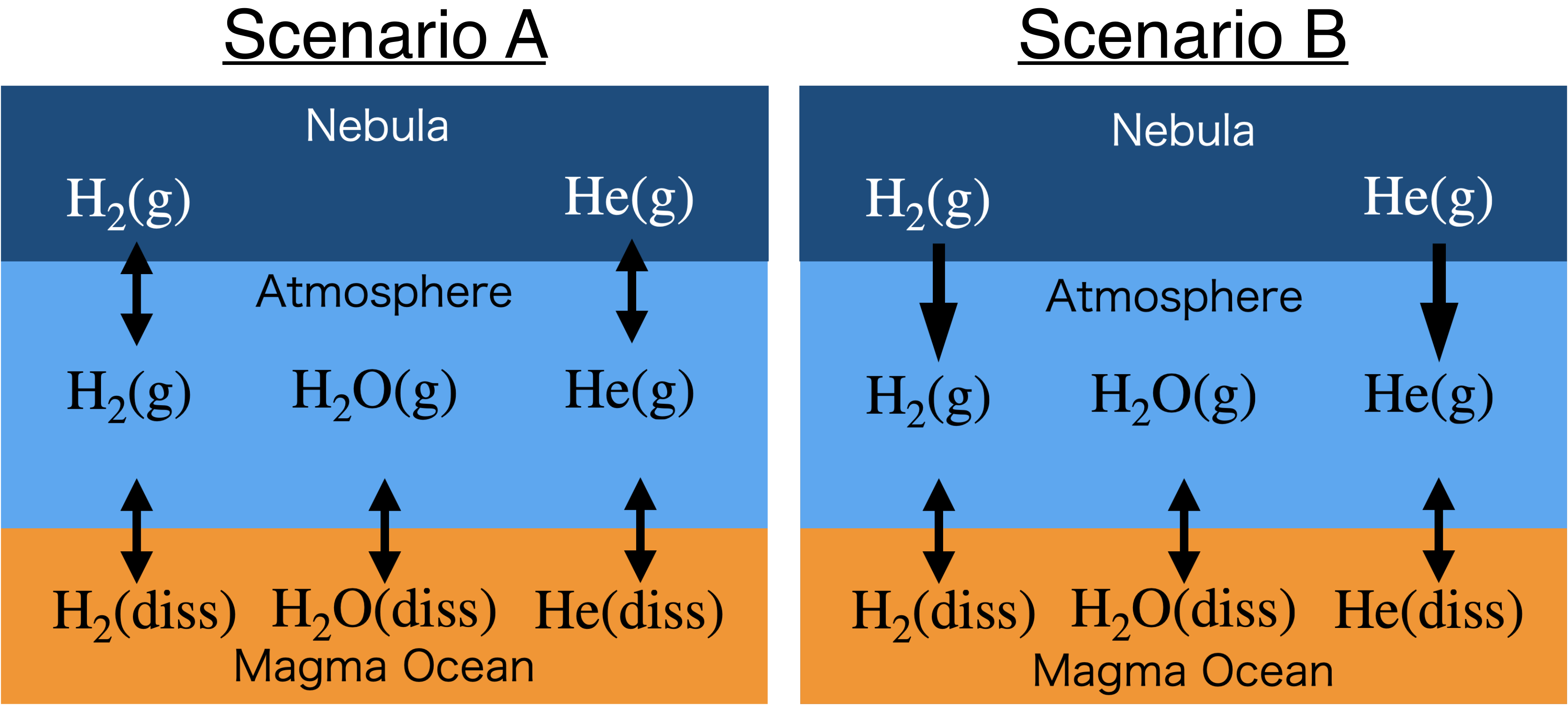}
    \caption{Two models adapted for equilibrium conditions between magma ocean, atmosphere, and disk gas in the formation stage (Section \ref{subsec:methods:workflow}). (A) case with an exchange of hydrogen and helium between the planetary atmosphere and disk gas. (B) case without exchange of hydrogen and helium between the planetary atmosphere and disk gas, where the atmosphere unilaterally accretes onto the planet.}
    \label{fig:scenarios}
\end{figure}

Scenario A assumes efficient exchange between the disk gas and the atmosphere during the atmosphere-magma ocean chemical reaction (\Cref{fig:scenarios}a). 
In Scenario A, we fixed the ratio of hydrogen to helium in the atmosphere to that of the disk gas, while water formed with the chemical reactions with the magma is added to the atmosphere. Scenario B assumes no disk-gas atmosphere exchange while the chemical reactions proceed (\Cref{fig:scenarios}b). 
In Scenario B, we assumed that the ratio of the total amount of hydrogen atoms (the sum of hydrogen atoms in molecular hydrogen and water) to that of helium in the entire planet equals the hydrogen-to-helium ratio in the disk gas.
Three-dimensional simulations of gas flow around a protoplanet embedded in a disk found active exchange between disk gas and the atmosphere inside the Bondi radius \citep[e.g.,][]{Ormel+2015,Fung+2015,Kurokawa+Tanigawa2018,Kuwahara+2019,Kuwahara+Kuroakwa2024A&A...682A..14K}; thus, Scenario A is considered to be more plausible and treated as the nominal scenario.

Through all calculations in this study, the following parameters are fixed. 
The core mass fraction is set at $32.5\ \rm{wt}\%$, equivalent to that of Earth. 
The mass fraction of helium in the disk gas is fixed at $28\ \rm{wt}\%$ \citep[the protosolar value,][]{Lodders2003}, and the remaining fraction is considered as hydrogen. 
The composition of magma is assumed to be that of Bulk Silicate Earth (BSE) \citep[Table 1 of][]{Schaefer2009}, which is employed for the conversion of molar and mass fractions of \ce{FeO} in the magma ocean.

We note that this study focuses on analyzing the dependence on these parameters. A population-synthesis–type calculation with a denser parameter grid and a comparison with the observed exoplanet population are beyond the scope of this work and will be addressed in a future study.

\subsubsection{Calculation of the initial state}

To find a planetary structure which satisfies a given set of input parameters, we iteratively calculated the planetary structure by changing inputs of the structure model. The converged structure is hereafter called \textquotedblleft the initial state." 

First, by using the structural model (Section \ref{subsec:methods:structure}), we computed the atmospheric structure of the planet. The input parameters for this calculation include the orbital semi-major axis, planet mass, envelope mass fraction, mass fractions of hydrogen, water, and helium, and intrinsic luminosity. The calculations yield profiles of the temperature, pressure, radius, density, optical thickness, entropy structure of the planetary atmosphere, and the planetary radius. 
We adjusted the planetary luminosity until the obtained magma-atmosphere boundary temperature matched the given value.

Second, we calculated the chemical equilibrium between the planetary atmosphere and interior (Sections \ref{subsec:methods:reactions} and \ref{subsec:methods:dissolution}).
The chemical-equilibrium calculation requires input parameters including temperature and pressure at the magma-atmosphere boundary, planetary mass, envelope mass fraction, and \ce{FeO} mass fraction in the melt. 
This calculation procedure follows the methodology outlined in \cite{kite2020atmosphere}. The goal of this calculation is to find the amount of remaining \ce{FeO} which satisfies conservation of oxygen. 
For a specific \ce{FeO} amount, the amounts of hydrogen and water in the atmosphere and in the magma ocean are uniquely determined. 
We iteratively changed the \ce{FeO} amount to find a consistent solution.

Finally, the obtained ratios of hydrogen and helium are compared with the assumed values through the calculations.
The determination of initial conditions involves iterative structural calculations and chemical-equilibrium calculations until both the magma-atmosphere boundary temperature and elemental abundance ratios match the assumed conditions.

Following \cite{kite2020exoplanet}, we assumed that all reactions between \ce{H2} and the magma ocean occur only in the initial stages and do not take place during the evolution. 
This assumption is reasonable because the added envelope mass is significant compared to the initial amount of \ce{FeO}, leading to the consumption of most of the internal \ce{FeO} in the early stages.

\subsubsection{Calculation of the temporal evolution}

\begin{figure}
    \centering
    \includegraphics[width=\linewidth]{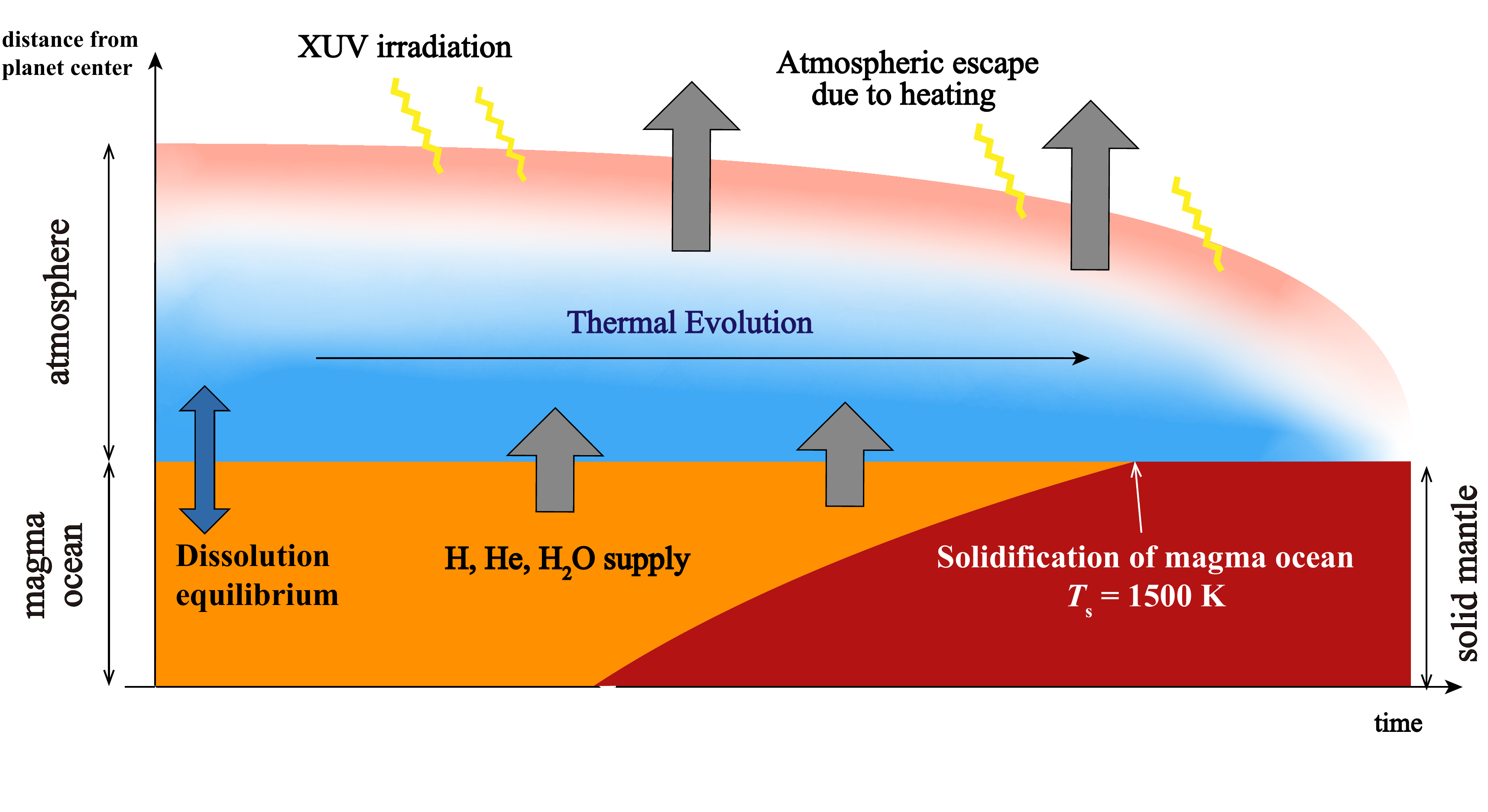}
    \caption{Schematic illustration of the planetary evolution in this study.}
    \label{fig:model}
\end{figure}

In the evolutionary calculation (\Cref{fig:model}), we considered the cooling (Section \ref{subsec:methods:cooling}) and atmospheric evolution with mass-loss (Section \ref{subsec:methods:escape}) and degassing (dissolution re-equilibration, Section \ref{subsec:methods:dissolution}). First, the planetary atmospheric structure was determined for each time step (Section \ref{subsec:methods:structure}). Using the temperature-entropy structure obtained from the structural calculations and the results from the previous time step, we checked whether the energy balance was satisfied (Equation \ref{eq:thermal}). We adjusted the luminosity and recalculated the structure until the energy balance was achieved.

Along with the results of the thermal evolution calculations, we also updated the planetary mass for each step.
We calculated the mass-loss rate (\Cref{eq:Mesc}) and updated the envelope mass. Iterative calculations were performed for structural and thermal evolution, updating the envelope mass to satisfy dissolution-equilibrium conditions, taking into account the total hydrogen, helium, and water masses of the planet.

Once the magma-atmosphere boundary temperature reaches $1500 \kel$, we assumed the magma ocean solidifies and the atmosphere-interior exchange halts. Although the solidification proceeds gradually from the bottom to the top as the magma ocean cools, we assumed an instant solidification at this temperature for the sake of simplicity.

\subsubsection{Termination conditions}

When $f_{\mathrm{env}} < 10^{-4}$, we stopped the calculation and treated all the atmosphere as being lost. 
In cases where the envelope mass did not reach the lower limit, we concluded the calculations at 5 billion years.

\section{Results}\label{sec:results}

In the following section, we first show the results of our evolution calculations with the nominal parameter set (Section \ref{subsec:results:nominal}). Next, we present the results of parameter survey for various planetary masses and orbital radii in the nominal case (Section \ref{subsec:results:mass_radius}). Then we change the values of input parameters one by one from the nominal case to elucidate the dependence on each parameter (Sections \ref{subsec:results:ox}--\ref{subsec:results:scenarios}). Finally, we compile our results to investigate whether the transition from the primary to the secondary atmospheres show any correlation with observables of sub-Neptunes (Section \ref{subsec:results:correlations}).

\subsection{Evolution in the nominal case}\label{subsec:results:nominal}

\begin{figure*}
    \centering
    \includegraphics[width=\linewidth]{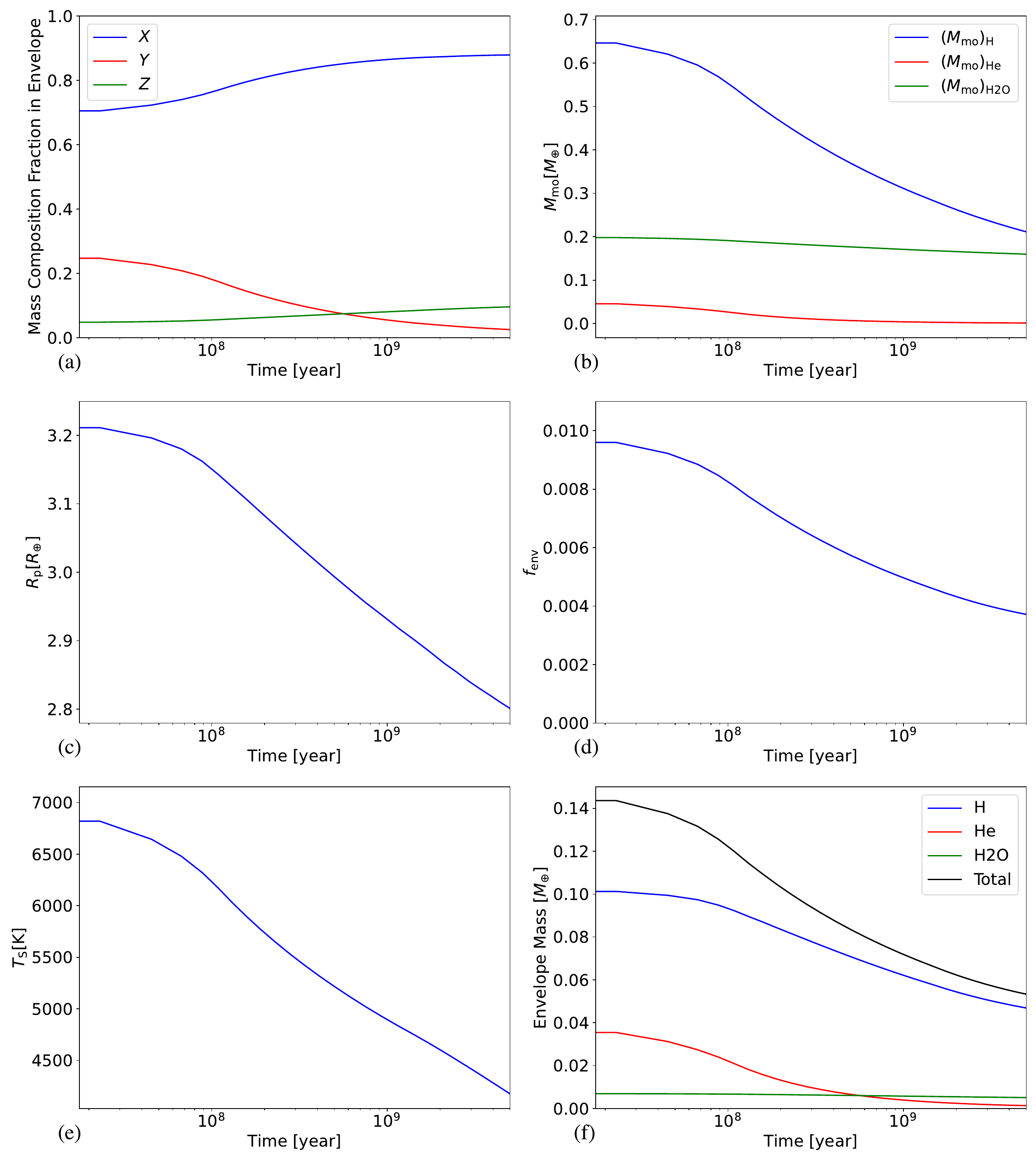}
    \caption{Evolution of a planet with the nominal parameter set (Table \ref{table: parameters}). Panels indicate (a) the mass fractions of hydrogen, helium, and water in the planetary atmosphere (denoted as $X$, $Y$, and $Z$), (b) the amounts of hydrogen, helium, and water in the magma ocean, (c) the planetary radius, (d) the envelope mass fraction (envelope mass/planetary mass), (e) the temperature at the magma-atmosphere boundary, and (f) the envelope mass.}
    \label{fig:evolution_nominal}
\end{figure*}

Evolution of a planet in the nominal case (Table \ref{table: parameters}) shows that atmospheric escape induces the transition from primary to secondary atmospheres (Figure \ref{fig:evolution_nominal}). 
Over time, the mass fraction of helium in the atmosphere significantly decreased, which leads to the formation of a helium-depleted atmosphere after 5 Gyrs (Figure \ref{fig:evolution_nominal}a). 
This change is attributed to atmospheric escape due to the XUV irradiation from the host star and the resultant replacement of the atmosphere. 
Atmospheric escape results in the decline in the envelope mass (Figure \ref{fig:evolution_nominal}d) and, consequently, in the magma-surface pressure, causing dissolved components to be released into the atmosphere (Figure \ref{fig:evolution_nominal}b). 
Because of helium solubility being lower than those of hydrogen and water, its initial amount in the interior is about 11 and four times lower than those of hydrogen and water, respectively. 
As a result, helium replenishment via degassing is limited, and the atmosphere becomes depleted in helium.

The atmospheric replacement can also be characterized by the increase in the relative mass fraction of water in the envelope. After 5 Gyrs, water surpassed helium to become the second most abundant gas in the envelope (Figure \ref{fig:evolution_nominal}a).
The absolute mass of water in the envelope was kept nearly constant with time, while those of hydrogen and helium declined (Figure \ref{fig:evolution_nominal}f).
This is attributed to the different pressure dependencies of the solubilities between water and the other two; in our model, the solubilities of hydrogen and helium are proportional to $p_i^1$ (Equations (\ref{eq: H2dissolution}) and (\ref{eq: Hedissolution})), whereas that of water is proportional to $p_\mathrm{H_2O}^{0.74}$ (Equation (\ref{eq: water dissolution})). 

Atmospheric loss leads to significant decline in the temperature and radius of the planet over time (Figures \ref{fig:evolution_nominal}c and \ref{fig:evolution_nominal}e). 
The magma-atmosphere boundary temperature decreased by about $3000 \kel$ in 5 Gyrs. 
However, since the solidification temperature of the magma ocean is assumed to be $1500 \kel$, the surface of the magma ocean remained molten, which allows for ongoing interactions between the atmosphere and magma.
Additionally, the planetary radius decreased by approximately $0.5\ \roplus$ after 5 Gyrs. 
A comparison with the result of a simulation without atmospheric escape (Appendix \ref{ap:wo_escape}) shows that the decline of the temperature and planetary radius is attributed not to cooling but chiefly to the decrease in the envelope mass fraction, a result of atmospheric escape.

\subsection{Dependence on planetary masses and orbital radii}
\label{subsec:results:mass_radius}

\begin{figure*}
    \centering
    \includegraphics[width=0.8\linewidth]{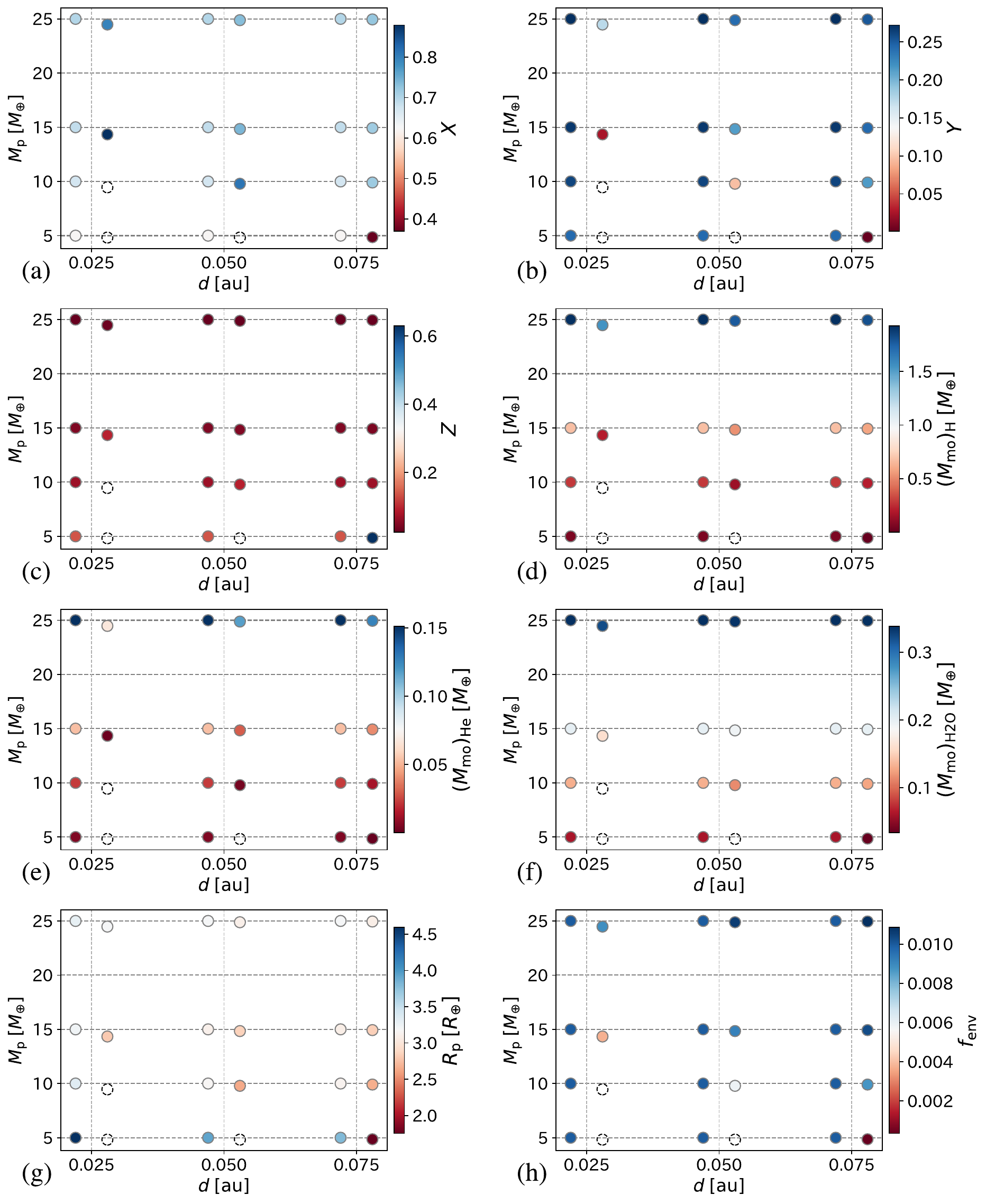}
    \caption{Dependence on planetary masses and orbital radii. The other parameters were assumed to be their nominal values (Table \ref{table: parameters}). (a)--(c) the mass fractions of hydrogen, helium, and water in the atmosphere, (d)--(f) the amounts of hydrogen, helium, and water in the magma ocean, (g) the planetary radius, and (h) and the envelope mass fraction. Vertical dashed lines correspond to the actual orbital distances ($d = 0.025$, $0.050$, and $0.075\ \mathrm{au}$) adopted in our calculations. For each $d$m the initial and final states are plotted with slight horizontal offsets to the left and right of the dashed lines, respectively, to distinguish them visually. Planets indicated with dashed circles did not retain their atmospheres over 5 Gyrs.}
    \label{fig:survey_nominal}
\end{figure*}

Next, we show the dependence of the planetary evolution on the masses and the orbital radius. The other parameters were set to be their nominal values (Table \ref{table: parameters}). In the following Figures \ref{fig:survey_nominal}–\ref{fig:survey_scenarios}, the vertical dashed lines indicate the orbital distances ($d = 0.025$, $0.050$, and $0.075\ \mathrm{au}$) at which the calculations were performed. For each $d$, the data points for the initial and final states are plotted with small horizontal offsets to the left and right of the dashed lines, respectively, to improve visibility. The offsets do not represent any difference in the parameter values.

We first focus on their initial states (data points offset to the left in Figures \ref{fig:survey_nominal}a--f).
Masses of hydrogen, helium, and water in their envelopes and interiors are dependent only on the planetary mass and not on the orbital radius (Figures \ref{fig:survey_nominal}a--f).
Larger planetary masses leads to higher pressures and thus enhanced dissolution in the magma ocean.
The mass fraction of water in the envelope decreases with increasing planetary mass (Figure \ref{fig:survey_nominal}c). 
As the planetary mass increases, more water dissolves in the interior, which reduces the relative amount of water in the atmosphere. 
In contrast, the change in the helium mass fraction was limited, because of its lower solubility (Figure \ref{fig:survey_nominal}b).
The lack of dependence on the orbital radius is due to the constant magma-atmosphere boundary temperature assumed across all masses and orbital radii. 

On the other hand, the planet radius is dependent on the distance from the host star (Figure \ref{fig:survey_nominal}g). 
Planetary atmospheres expand due to high temperatures close to the host star, which leads to larger planetary radii. 
This effect was particularly pronounced in lower-mass planets due to their lower gravitational potential energy (e.g., a planet with $5\moplus$ and 0.025 au in \Cref{fig:survey_nominal}g).

As planets evolved (data points offset to the right in Figure \ref{fig:survey_nominal}), helium-depleted secondary atmospheres formed in a parameter region with intense atmospheric escape. 
Planets underwent different changes in atmospheric composition and planet radius, depending on their masses and orbital radii. 
In cases where atmospheres were retained, planets closer to the star and with lower mass comprise atmospheres with higher hydrogen and water and lower helium mass fractions (Figures \ref{fig:survey_nominal}a--c).
Greater XUV flux and the reduced gravitational potential energy for these planets lead to higher atmospheric escape rates, and their atmospheres were replaced with the degassed components. 
This is also evident from the observed reduction in hydrogen and water contents in the magma ocean (Figures \ref{fig:survey_nominal}d and \ref{fig:survey_nominal}f).
The atmospheric loss also causes significant decrease in planetary radii in the same parameter region with the primary to secondary transition (Figure \ref{fig:survey_nominal}g).

Some low-mass planets close to their host stars were unable to retain their atmospheres after 5 Gyrs (cases marked with dashed circles in Figure \ref{fig:survey_nominal}). 
On the other hand, planets with higher masses and larger orbital radii retained their primary atmospheres. 
In some cases of these distant planets, the post-evolution envelope mass slightly increased from their initial state (Figure \ref{fig:survey_nominal}h). 
Since the solubility of hydrogen in the magma ocean is temperature-dependent (Equation (\ref{eq: H2dissolution})), a decrease in temperature due to thermal evolution leads to the release of dissolved hydrogen (Section \ref{subsec:results:nominal}).

As shown in Section \ref{subsec:results:nominal}, the hydrogen mass fraction generally increases as atmospheric loss proceeds (Figure \ref{fig:survey_nominal}a), while the proportion of water in the atmosphere remains at a few to ten wt\% (Figure \ref{fig:survey_nominal}c). 
The only exception is seen in the result for $5\moplus$ at $0.075 \, \mathrm{au}$, where approximately 60 wt\% of the final atmosphere consists of water, with the remaining 40\% being mostly hydrogen. 
This is due to the very low final envelope mass fraction of the planet; the envelope mass fraction is about 0.1 wt\% (Figure \ref{fig:survey_nominal}h). The planet radius falls below 2 Earth radii (Figure \ref{fig:survey_nominal}g). 
This indicates the replacement to the secondary atmosphere with little helium due to escape and degassing from the interior. 
The reason why water exceeds hydrogen in the mass fraction is attributed to the different pressure-dependencies of their solubilities in magma (Equations (\ref{eq: H2dissolution}) and (\ref{eq: water dissolution})).

\subsection{Dependence on the oxidation state of magma ocean}
\label{subsec:results:ox}

\begin{figure}
    \centering
    \includegraphics[width=\linewidth]{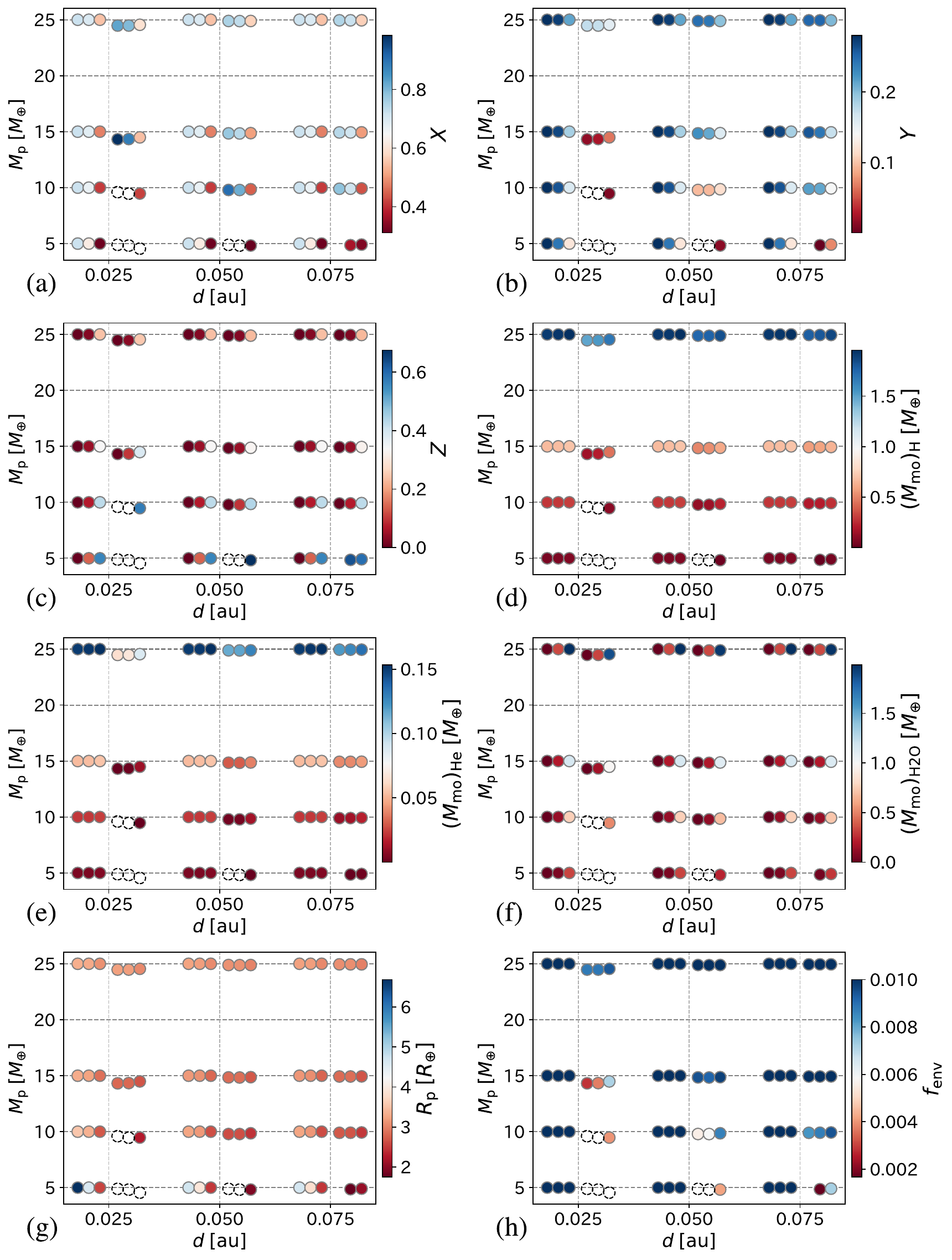}
    \caption{Same as Figure \ref{fig:survey_nominal} but showing the dependence on the oxidation state of magma ocean. Three successive data points corresponds to $\wfeo = 0\, \mathrm{wt\%}$, $\wfeo = 8.24\, \mathrm{wt\%}$ (nominal), and $\wfeo = 49\, \mathrm{wt\%}$, from left to right. The other parameters were assumed to be their nominal values (Table \ref{table: parameters}) We note that missing data points indicate where the calculation failed to obtain a solution in the process of evolution.}
    \label{fig:survey_redox}
\end{figure}

Planets with more oxidizing magma oceans initially possess larger amounts of water in their atmospheres and interiors. 
Figure \ref{fig:survey_redox} compares the results for planets with reducing ($\wfeo = 0 \, \mathrm{wt}\%$), nominal ($\wfeo = 8.24 \, \mathrm{wt}\%$), and oxidizing ($\wfeo = 49 \, \mathrm{wt}\%$) magma oceans. 
Again, we first focus on their initial states (data points offset to the left in Figure \ref{fig:survey_redox}).
The water mass fraction in the envelope (Figure \ref{fig:survey_redox}c) and the amount of water in the interior (Figure \ref{fig:survey_redox}f) becomes higher as $\wfeo$ increases. 
Water dilutes hydrogen (Figure \ref{fig:survey_redox}a) and helium (Figure \ref{fig:survey_redox}b) from their disk gas values ($X=0.72$ and $Y=0.28$), and also makes the planetary radius smaller (Figure \ref{fig:survey_redox}g).

Planets with higher water amounts are less susceptible to escape, and thus higher $\wfeo$ values leads to extended lifetime and the retention of atmospheres over wider parameter space (data points offset to the right in Figure \ref{fig:survey_redox}). 
Higher water production leads to a smaller planetary radius and, consequently, a reduced atmospheric escape rate, thereby extending the parameter space for the atmospheric retention (Figure \ref{fig:survey_redox}h). 
While initial water production dilutes helium in the atmosphere, a reduced atmospheric escape rate allows the retention of primordial helium for the parameter space where the transition to the secondary atmosphere was observed for reducing cases (e.g., the case with $d = 0.075\ \mathrm{au}$ and $M_\mathrm{p} = 5\ M_\oplus$ in Figure \ref{fig:survey_redox}b).

\subsection{Dependence on the initial envelope mass fraction}\label{subsec:results:fenv}

\begin{figure}
    \centering
    \includegraphics[width=\linewidth]{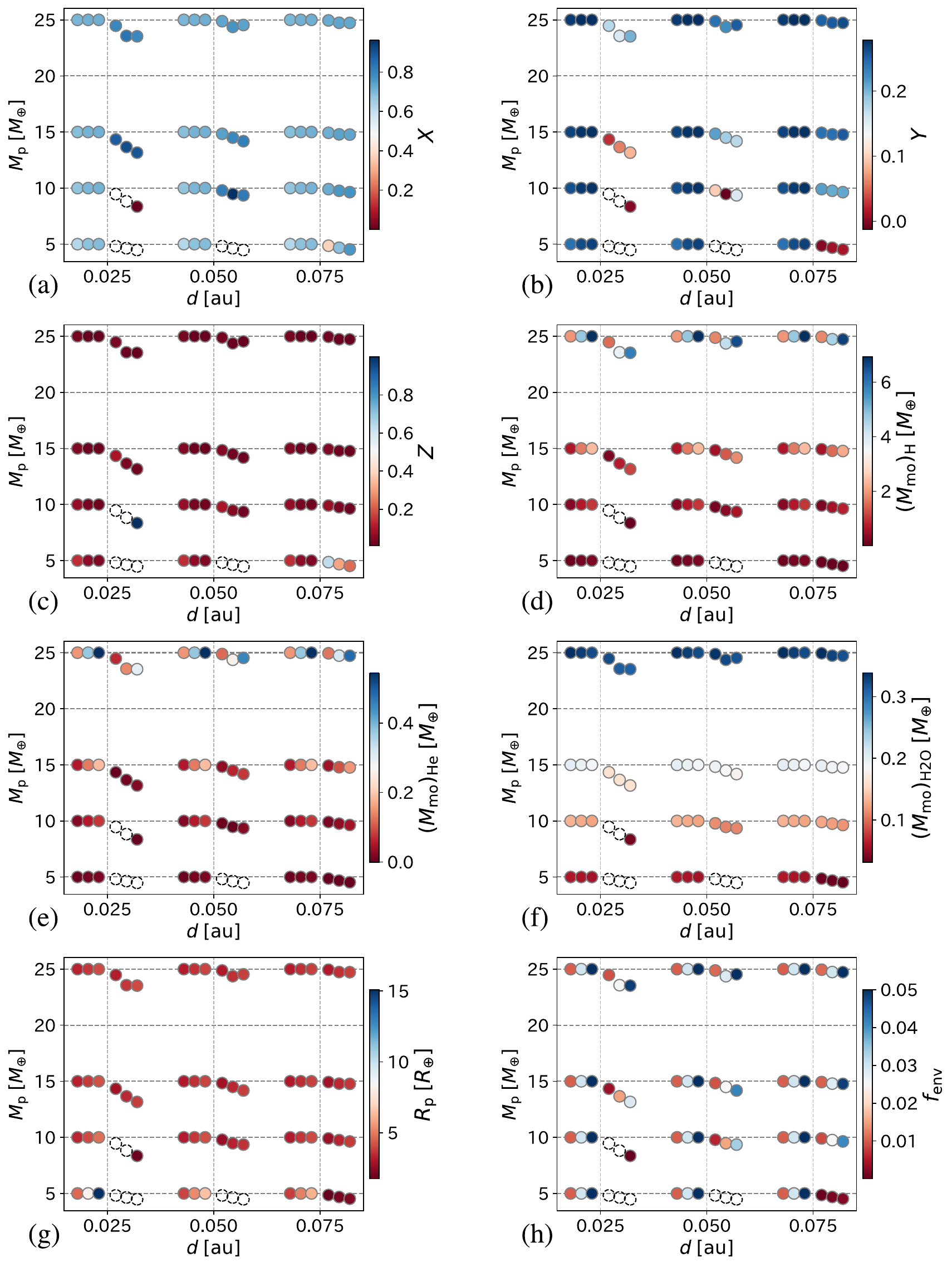}
    \caption{Same as Figure \ref{fig:survey_nominal} but showing the dependence on the initial envelope mass. Three successive data points corresponds to $\fenvini = 1\ \mathrm{wt \%}$ (nominal), $\fenvini = 3\ \mathrm{wt \%}$ , and $\fenvini = 5\ \mathrm{wt \%}$ from left to right. The other parameters were assumed to be their nominal values (Table \ref{table: parameters}).}
    \label{fig:survey_fenv}
\end{figure}

A higher initial envelope mass fraction leads to the formation of a more hydrogen-rich atmosphere (data points offset to the left in Figure \ref{fig:survey_fenv}). 
Planets with $\fenvini = 3\ \mathrm{wt \%}$ and $\fenvini = 5\ \mathrm{wt \%}$ tend to have higher hydrogen fractions than the nominal $\fenvini = 1\ \mathrm{wt \%}$ case (Figure \ref{fig:survey_fenv}a). 
This is because a higher atmospheric mass fraction leads to relative decrease in the amount of water produced (Figure \ref{fig:survey_fenv}c). 
As we assumed a fixed hydrogen to helium ratio for the atmosphere (Scenario A in Figure \ref{fig:scenarios}), the resultant helium mass fraction is higher as well (Figure \ref{fig:survey_fenv}b). 
Because of the higher envelope mass fraction and its lower mean-molecular-weight, the planetary radii are larger compared to the nominal cases (Figure \ref{fig:survey_fenv}g).

As planets evolve (data points offset to the right in Figure \ref{fig:survey_fenv}), a higher envelope mass fraction allows retention of the atmosphere for wider parameter space.
As their initial atmospheres were massive and rich in hydrogen, the compositional change due to atmospheric escape and replenishment is limited in the $\fenvini = 5\ \mathrm{wt \%}$ case compared to the nominal $\fenvini = 1\ \mathrm{wt \%}$ case (Figures \ref{fig:survey_fenv}a--c). 
Their planetary radii remains larger as well (Figure \ref{fig:survey_fenv}g), thanks to the higher envelope mass fraction and the high proportion of hydrogen in the envelope retained. 
Having said that, the amounts of helium decreased in regions with high atmospheric escape rates, suggesting the replacement to secondary atmospheres (Figure \ref{fig:survey_fenv}b).

\subsection{Dependence on the initial temperature}
\label{subsec:results:temperature}

The initial temperature did not change our results significantly (figures not presented here as they look similar to Figure \ref{fig:survey_nominal}).
The initial conditions including atmospheric compositions, dissolved contents, and planetary radii are similar with those in the nominal case. 
Slight differences include the amounts of dissolved hydrogen and helium being slightly lower for $T_{\rm s} = 5000\, \mathrm{K}$ due to the temperature-dependence of their solubilities (Equations (\ref{eq: H2dissolution}) and (\ref{eq: Hedissolution})).
Additionally, the planets' radii were smaller for $T_{\rm s} = 5000\, \mathrm{K}$.
However, these differences were not substantial enough to cause significant changes in planetary evolution.

\subsection{Dependence on the scenario for gas accretion and chemical reactions}\label{subsec:results:scenarios}

\begin{figure}
    \centering
    \includegraphics[width=\linewidth]{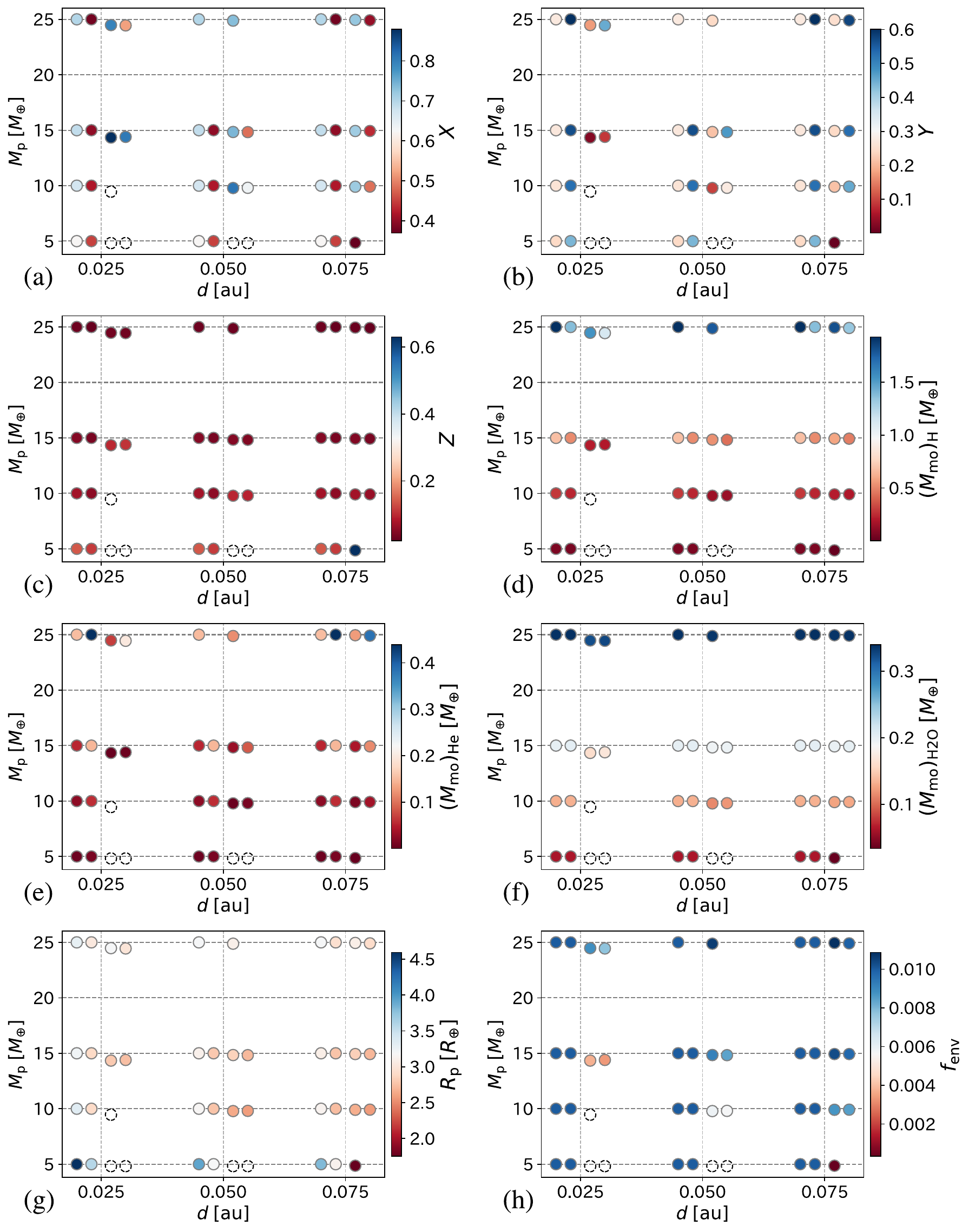}
    \caption{Same as Figure \ref{fig:survey_nominal} but showing the dependence on the accretion scenario (Figure \ref{fig:scenarios}). Two successive data points corresponds to Scenarios A (nominal) and B from left to right. The other parameters were assumed to be their nominal values (Table \ref{table: parameters}). We note that missing data points indicate where the calculation failed to obtain a solution in the process of finding the initial state or of evolution.}
    \label{fig:survey_scenarios}
\end{figure}

In another scenario where gas accretes unilaterally from the disk to a planet (Scenario B; Section \ref{subsec:methods:workflow} and Figure \ref{fig:scenarios}), planets form with a higher helium mass fraction in their atmospheres (data points offset to the left in Figure \ref{fig:survey_scenarios}). 
The initial mass fractions of helium in their atmospheres range 40--60 wt\%, which are significantly higher than those in Scenario A (Figure \ref{fig:survey_scenarios}b). 
While hydrogen and water dissolve efficiently in the planetary interior, helium is less soluble and thus remains in the atmosphere. 
As the bulk hydrogen to helium ratio was fixed in Scenario B, the initial helium fraction in the atmosphere became higher in Scenario B than that in Scenario A.

Planets in Scenario B evolve with higher helium mass fractions in their atmospheres than those in Scenario A (data points offset to the right in Figure \ref{fig:survey_scenarios}). 
Except for planets with $5$, $10$, and $15\moplus$ at $0.025\ {\rm au}$, helium fractions higher than the disk gas value ($Y=0.28$) were maintained (Figure \ref{fig:survey_scenarios}b). 
This is because planets less affected by atmospheric escape tend to retain their initial atmospheric compositions. 
On the other hand, in regions with high rates of atmospheric escape ($5$, $10$, and $15\ \moplus$ at $0.025\ {\rm au}$), planets evolved to sub-solar helium mass fractions.

\subsection{Relations between the planetary radius, atmospheric composition, and escape rate}\label{subsec:results:correlations}

\begin{figure*}
    \centering
    \includegraphics[width=\linewidth]{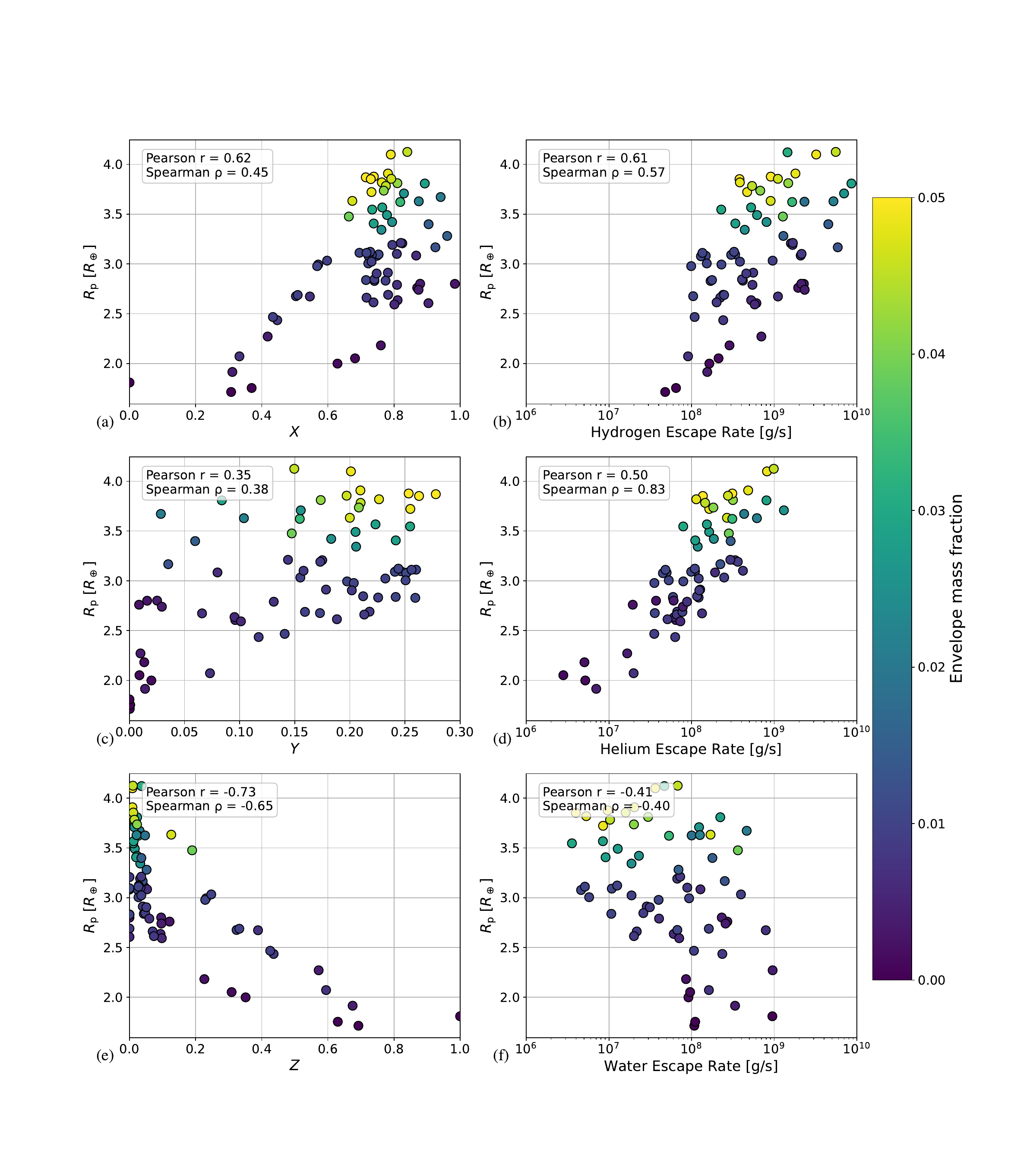}
    \caption{Relations between the planetary radius, atmospheric composition, and escape rate. Plotted are all the final states obtained from our parameter survey, except for those in Scenario B. Panels (a)--(f) show hydrogen, helium, water mass fractions in the envelope and their escape rates. The data points are color-coded by the envelope mass fraction. The correlation coefficients (Pearson $r$ and Spearman $\rho$) are shown in each panel.}
    \label{fig:survey_all}
\end{figure*}

A compilation of all results shows correlations between the planet radius and both atmospheric composition and escape rates (Figure \ref{fig:survey_all}).
The planetary radius shows moderate to weak positive correlations with hydrogen and helium mass fraction (Spearman $\rho=0.45$ and $0.38$, respectively), and a negative correlation with water fraction (Spearman $\rho=-0.65$) (Figures \ref{fig:survey_all}a, \ref{fig:survey_all}c, and \ref{fig:survey_all}e). 
We found a strong positive (Spearman $\rho=0.83$) correlation between the planetary size and the escape rate of helium, which can be constrained from transit observations in the helium triplet line (Figure \ref{fig:survey_all}d).
This is caused by two factors.
One is the weak correlation between the helium abundance and the planetary radius mentioned above (Figure \ref{fig:survey_all}c).
Another is the positive correlation between the escape rate and the planetary radius; larger planets tend to be less dense, experiencing higher escape rates (Equation (\ref{eq:Mesc})), which in turn reinforces the correlation between the helium escape rate and the planetary radius.
Correlations with the planetary radius are less clear (Spearman $|\rho|$-values are lower) for escape rates of hydrogen and water (Figures \ref{fig:survey_all}b and \ref{fig:survey_all}f).
We note that, although our model does not treat photolysis, water will escape not as molecular H$_2$O but as dissociated/ionized forms in reality, which may further obscure the correlation for the hydrogen escape rate.

To further illustrate what controls the distribution, we examined alternative versions of the figure in which the data points are color-coded by the envelope mass fraction, the planetary mass, and the orbital distance (Figures \ref{fig:survey_all}, \ref{fig:survey_all_mass}, and \ref{fig:survey_all_distance}; the latter two are provided in Appendix \ref{ap:figures}). These comparisons indicate that no single parameter dictates the overall trend. Instead, lower remaining envelope mass fractions (and thus smaller initial ones), smaller planetary masses, and shorter orbital distances collectively result in smaller planetary radii, lower hydrogen and helium abundances (and their escape rates), and higher water abundances (and escape rates). These relationships are primarily driven by atmospheric escape, as evidenced by the model run without escape, which shows little change in planetary radius and composition (Appendix \ref{ap:wo_escape}). Atmospheric escape and replenishment together shaped the population of sub-Neptunes. Planets with secondary atmospheres (poor in hydrogen and helium and rich in water) have smaller sizes; typically, planets whose sizes are $\lesssim 2.5\ R_\oplus$ show the helium mass fraction $Y<0.14$, a half of the protosolar value (Figure \ref{fig:survey_all}c).

\section{Discussion}\label{sec:discussion}

\subsection{Comparison with the observations of escaping helium} \label{subsec:discussion:observations}

\begin{figure}
    \centering
    \includegraphics[width=\linewidth]{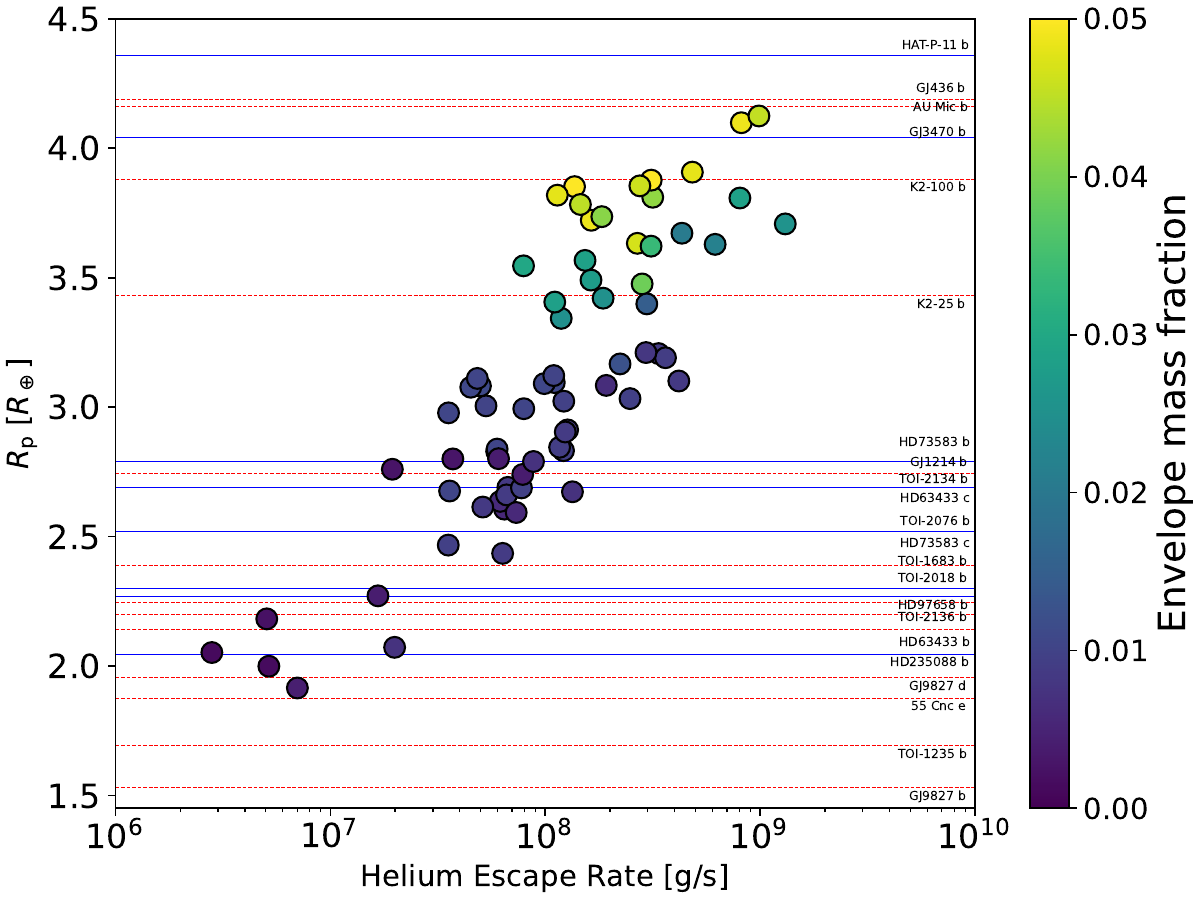}
    \caption{Planets with reported detection or non-detection of escaping helium overlaid with the model results in Figure \ref{fig:survey_all}d. Solid-blue and dashed-red lines indicate detection and non-detection, respectively. Observation data are from \cite{Orell-Miquel+2024A&A...689A.179O}.}
    \label{fig:survey_all_obs}
\end{figure}

Although detailed comparison to observations of the helium triplet absorption is beyond the scope of this study, here we discuss whether the trend shown in our model results is relevant to helium detection/non-detection qualitatively. 
Figure \ref{fig:survey_all_obs} shows a compilation of helium observations \citep{Orell-Miquel+2024A&A...689A.179O}, overlaid with our planet size-helium escape rate diagram.
While our results show a positive correlation between the planet size and helium escape rate, observational detection and non-detection do not simply correlate with the planet size; this supports that the transit depth of the helium triplet line is controlled by several factors including stellar properties \citep{Krishnamurthy+Cowan2024AJ....168...30K,Sanz-Forcada+2025A&A...693A.285S}.
However, we suggest that the transition to the secondary atmosphere (the depletion of helium) may account for non-detection of escaping helium from small-sized planets (typically $\lesssim 2.5\ R_\oplus$).

We note that our comparison does not include factors such as stellar properties, age, planet mass, and orbital semi-major axis and thus should be considered with caution. 
All results in this study are based on calculations using the bolometric and XUV luminosities of sun-like stars. 
The absorption of the escaping helium triplet from planetary atmospheres is not solely determined by the amount of helium escaping from planets. 
Theory predicted that high XUV and low mid-UV fluxes are preferred for populating helium in the metastable triplet state \citep{oklop2019}, which was later supported by observations \citep{Krishnamurthy+Cowan2024AJ....168...30K}. 
Moreover, as these observations are conducted from Earth, the direction and size of the region where the escaping helium spreads affects the absorption \citep{rumenskikh2023}. 
A more detailed comparison, incorporating these additional conditions is needed in a future study.

\subsection{Effects of escape-induced fractionation and other escape mechanisms} \label{subsec:discussion:fractionation}

As noted in Section \ref{subsec:methods:escape}, our model does not take escape-induced elemental fractionation into account.
A low hydrogen escape rate can leave helium behind, leading to helium enrichment as the escape proceeds \citep{hu2015,cherubim2024Strong}, which may counterbalance the depletion of helium caused by atmospheric replacement.
The escape-induced fractionation is expected to influence only a limited region within our parameter space (see Section \ref{subsec:methods:escape}).
However, about half of the small-sized ($\lesssim 2.5\ R_\oplus$) planets discussed in Section \ref{subsec:discussion:observations} reside at orbits with periods $\gtrsim 5\ \mathrm{days}$ \citep[see Figure 10 of][]{Orell-Miquel+2024A&A...689A.179O}, where the fractionation effect may become important \citep{cherubim2024Strong}.
A future study is desired to address the interplay between escape-induced fractionation (helium enrichment) and atmospheric replacement (helium depletion), as well as its implications for the observed detection and non-detection of helium.

Although we considered only the XUV-driven escape as the mechanism of atmospheric mass-loss, other mechanisms such as boil-off \citep{Owen&Wu2016}, core-powered mass-loss \citep{Ginzburg2018}, and impact erosion \citep{Izidoro2022} can drive atmospheric replacement. These additional mass-loss mechanisms may allow the transition to secondary atmospheres on larger-size planets than we predicted.

\subsection{Evaporation efficiency and XUV luminosity evolution}

Although we assumed a fixed evaporation efficiency ($\eta = 0.1$) and a specific XUV luminosity model \citep{ribas2005}, both parameters can vary in reality. Hydrodynamic simulations of atmospheric escape indicate variations by a factor of a few \citep{owen&jackson2012,Caldiroli+2021A&A...655A..30C,Caldiroli+2022A&A...663A.122C}. Molecular-kinetic modeling suggests that, in the transonic escape regime, the escape rate no longer increases with the incident XUV flux \citep{Johnson+2013ApJ...768L...4J}. Moreover, the XUV luminosities of young stars exhibit large scatter even among stars of similar ages \citep{Sanz-Forcada+2011A&A...532A...6S,Tu+2015A&A...577L...3T}. Adopting different prescriptions for these parameters would shift the quantitative locations of the transition boundaries, but the qualitative trend of helium depletion is expected to remain robust.

\subsection{Limitations of solubility and chemical equilibrium models}

The equilibrium chemistry models used in this study are based on experimental data from the field of geochemistry, but we note that the available experimental values are limited for parameter spaces relevant for hot sub-Neptunes.
More lab and numerical experiment data for material properties at the temperature and pressure range of magma-atmosphere interfaces on sub-Neptunes are critically needed.

For instance, the Gibbs free energy for Reaction (\ref{reaction:H2FeO}) was evaluated by extrapolation beyond the experimentally constrained temperature range. This uncertainty may influence the estimated amount of water produced for a given initial FeO content. Nevertheless, because a wide range of FeO contents was explored, the qualitative trend obtained in this study is considered to be robust.

The upper-pressure limit for the water solubility model used in this study is 1 GPa \citep{Papale1997}.
However, for an envelope mass fraction of 5 wt\%, the planetary surface pressure can be as high as 10 GPa.
In reality, the upper limit of sub-Neptunes' envelope mass fraction is several times higher, and so the pressure at the magmatic-atmospheric boundary could be up to several tens of GPa.
We still used the solubility equation (\Cref{eq: water dissolution}) in these high-pressure regions, but this may affect the results.

While the solubility models used in this study are determined by pressure, in reality, fugacity is expected to dictate solubility. 
However, to the best of our knowledge, no models formalizing the solubility of hydrogen or water based on fugacity currently exist at sub-Neptune conditions. 

\section{Conclusions}\label{sec:conclusions}
This study aimed to elucidate how short-period sub-Neptunes with hydrogen, helium, and water vapor atmospheres evolve with time via combination of XUV-driven atmospheric escape and replenishment from their interiors. 
Using a 1D structure model, we conducted evolutionary calculations, considering water production, XUV-driven atmospheric escape, and degassing from the planetary interior.
We found the transition from primary atmospheres derived from disk gas accretion to secondary atmospheres supplied by degassing for highly-irradiated, low-mass planets.
The helium abundance in the atmosphere as well as its escape rate were shown to serve as proxies for the transition.
Water production, depending on the amount of \ce{FeO} in the magma ocean assumed, extended the lifetime of the planetary atmosphere.
The extent to which elemental exchange occurs between disk gas and atmosphere during atmospheric formation (accretion scenarios) affects the atmospheric composition significantly, whereas efficient exchange that leads to a fixed hydrogen to helium ratio was considered to be more realistic. 
Our model predicts a correlation between planetary radius and the helium abundance in the atmosphere after several Gyr evolution, for a range of input parameters including the distance from the host star, planetary mass, mantle FeO content, and the initial envelope mass. 
Moreover, we found a trend where the atmospheric escape rate of helium, which can potentially be constrained from transit observations in the metastable helium triplet line, increases with the planetary radius; smaller planets are less likely to have detectable escaping helium due to the transition to secondary atmospheres. 
This may account for non-detection of helium absorption for small-size planets (typically $\lesssim 2.5\ R_\oplus$).  Future work is needed for more detailed comparison considering other factors such as stellar properties.

\section*{acknowledgments}
We thank Dr. Kenji Kurosaki for providing the opacity table of water. This study was supported by JSPS KAKENHI Grant Number 20KK0080, 22H01290, 22H05150, 21H04514, 23K22561.

%






\appendix

\restartappendixnumbering

\section{Evolution without atmospheric escape}
\label{ap:wo_escape}

\begin{figure}
    \centering
    \includegraphics[width=\linewidth]{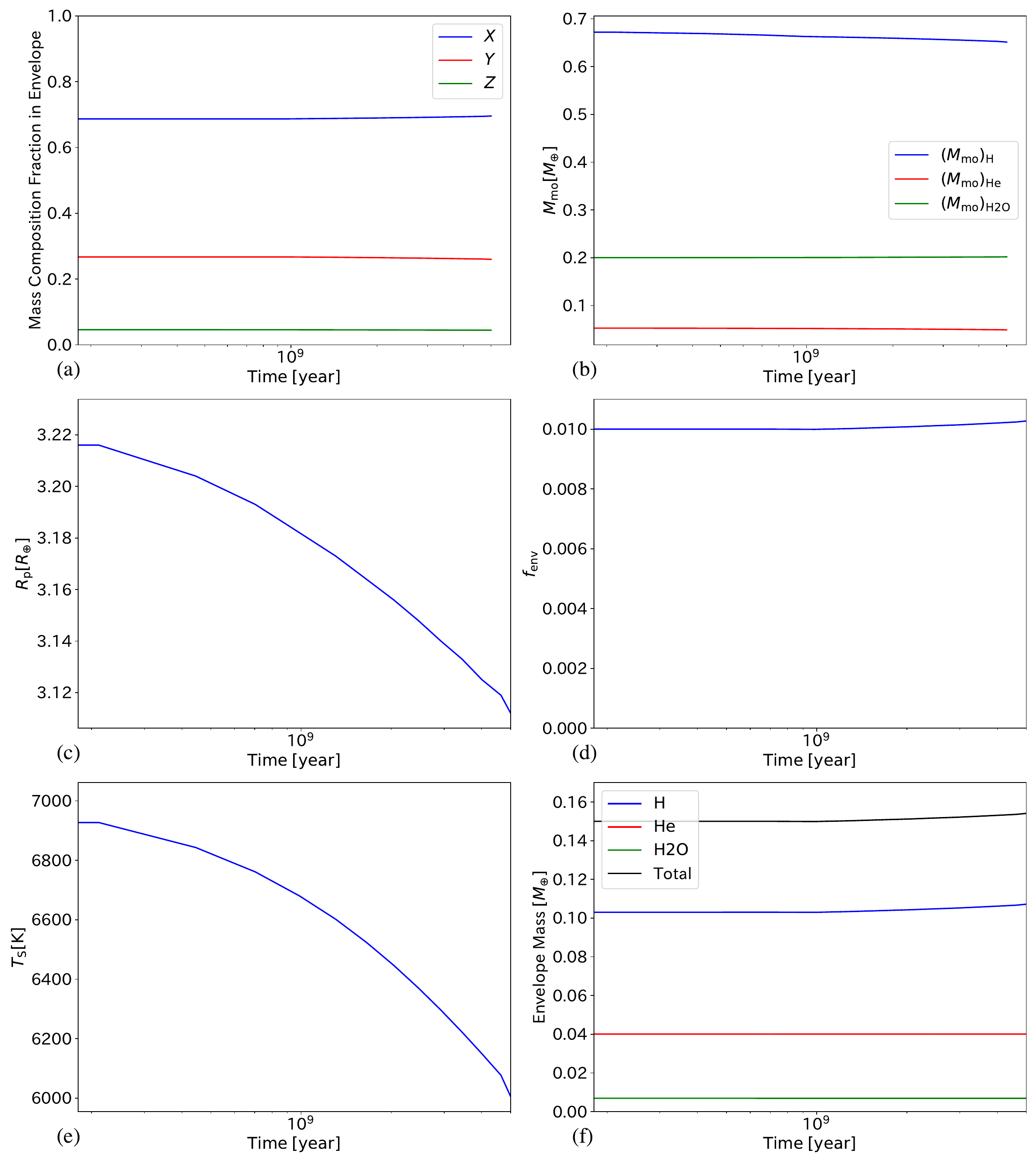}
    \caption{Evolution of a planet without atmospheric escape (thermal evolution only). The input parameters follows the nominal values (Table \ref{table: parameters}). Panels indicate (a) the mass fractions of hydrogen, helium, and water in the planetary atmosphere (denoted as $X$, $Y$, and $Z$), (b) the amounts of hydrogen, helium, and water in the magma ocean, (c) the planetary radius, (d) the envelope mass fraction (envelope mass/planetary mass), (e) the temperature at the magma-atmosphere boundary, and (f) the envelope mass.}
    \label{fig:evolution_coolingonly}
\end{figure}

To ascertain the impact of atmospheric escape on the evolution of sub-Neptunes, we conducted calculations without atmospheric escape for reference. 
The results presented in this section considered water production, dissolution into the planetary interior, and thermal evolution, but did not take into account the atmospheric escape evolution.
The input parameters followed the nominal values (Table \ref{table: parameters}).

Thermal evolution reduced the magma-atmosphere boundary temperature, leading to a slight but noticeable release of elements dissolved in the interior into the atmosphere. 
Figure \ref{fig:evolution_coolingonly} shows the results without atmospheric escape. 
In 5 Gyr evolution, the surface temperature dropped by about 1000 K (Figure \ref{fig:evolution_coolingonly}e), which led to decrease in the planetary radius by a few \% (Figure \ref{fig:evolution_coolingonly}c). 
While the atmospheric composition remained largely unchanged over 5 billion years (Figure \ref{fig:evolution_coolingonly}a), there was slight increase in atmospheric \ce{H2} (Figure \ref{fig:evolution_coolingonly}f), attributed to degassing (Figure \ref{fig:evolution_coolingonly}b) caused by its reduced solubility due to the temperature decline. 
However, degassing due to thermal evolution only was minimal, contributing increase in the envelope mass by only a few \% (Figure \ref{fig:evolution_coolingonly}d). 
The interior retains about four times more hydrogen than the atmosphere, with most of it remaining within the planet after 5 billion years (Figures \ref{fig:evolution_coolingonly}b and \ref{fig:evolution_coolingonly}f). 

Helium, dissolved in the interior in quantities similar to the atmosphere, did not undergo significant degassing to change the atmospheric composition (Figures \ref{fig:evolution_coolingonly}b and \ref{fig:evolution_coolingonly}f).
The amount of water constitutes about 5 wt\% of the atmosphere (Figures \ref{fig:evolution_coolingonly}f). 
A substantial amount of water, more than 20 times the atmospheric water content, is dissolved in the interior; however, again its degassing was minimal and did not influence the atmospheric composition (Figure \ref{fig:evolution_coolingonly}b).

\restartappendixnumbering

\section{Additional Figures}
\label{ap:figures}

\begin{figure}
    \centering
    \includegraphics[width=\linewidth]{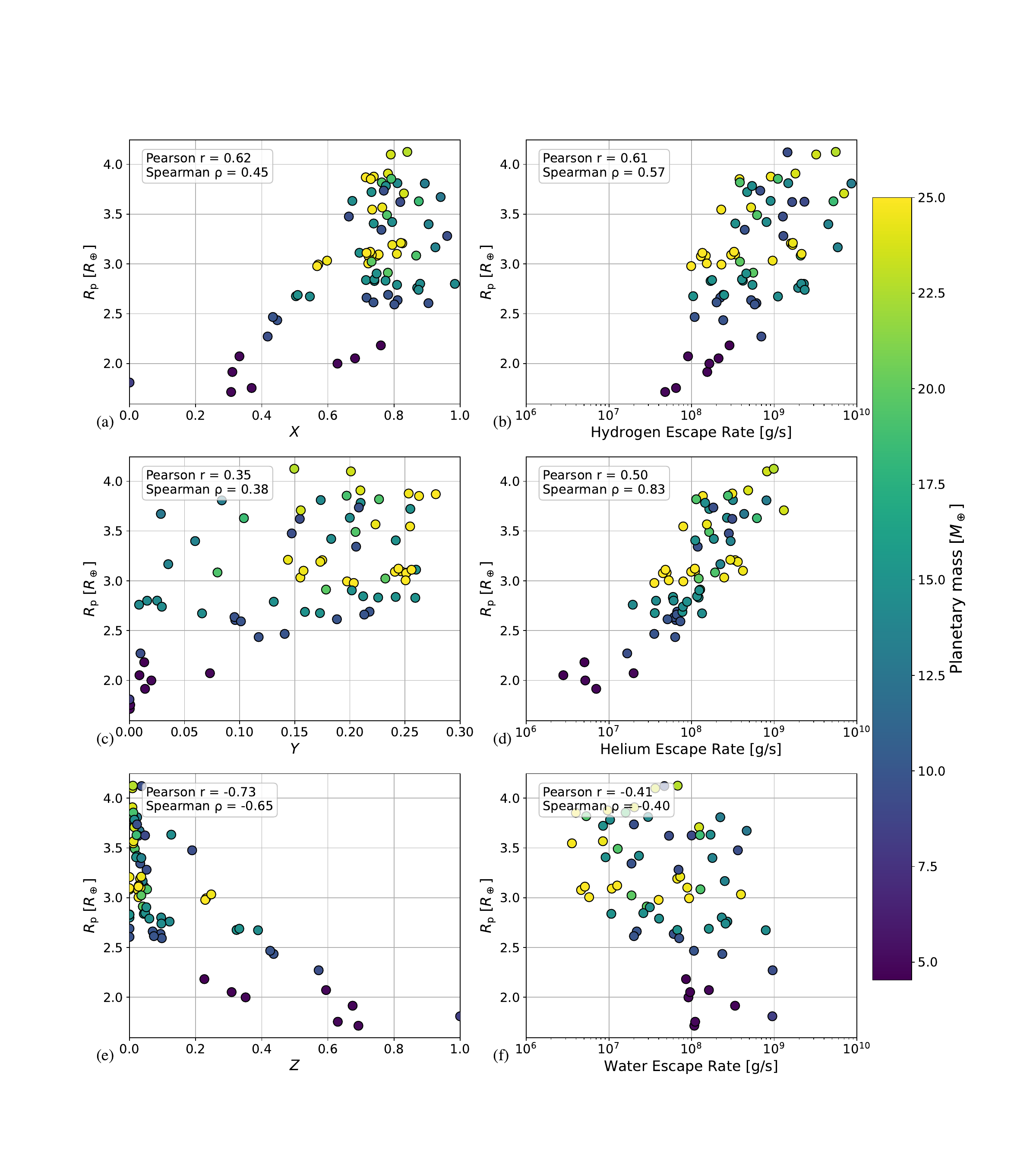}
    \caption{Same with Figure \ref{fig:survey_all} but color-coded by the planetary mass.}
    \label{fig:survey_all_mass}
\end{figure}

\begin{figure}
    \centering
    \includegraphics[width=\linewidth]{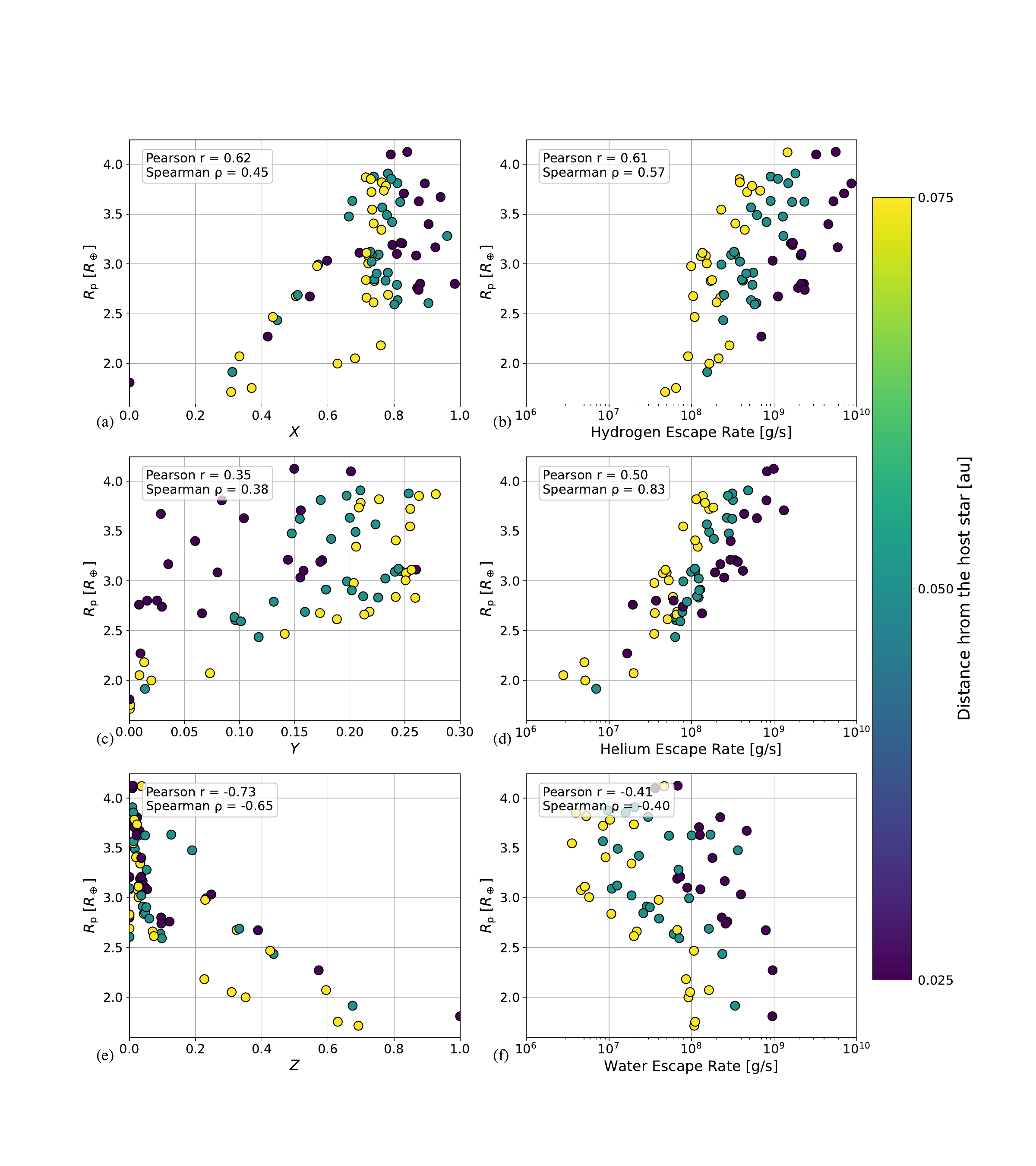}
    \caption{Same with Figure \ref{fig:survey_all} but color-coded by the distance from the host star.}
    \label{fig:survey_all_distance}
\end{figure}

Figures \ref{fig:survey_all_mass} and \ref{fig:survey_all_distance} present the distributions of the final atmospheric compositions and escape rates, color-coded by planetary mass and orbital distance.


\bibliography{bibliography}{}
\bibliographystyle{aasjournal}



\end{document}